\documentclass[]{emulateapj}

\def\lya{Ly$\alpha$ }
\def\Msun{M_\odot}
\def\hMsun{h^{-1}M_\odot}
\def\hMpc{h^{-1}{\rm Mpc}}
\def\hkpc{h^{-1}{\rm kpc}}
\def\VhMpc{h^{-3}{\rm Mpc}^3}
\def\kms{{\rm km\, s^{-1}}}
\def\Lint{L_{\rm intrinsic}}
\def\Lapp{L_{\rm apparent}}
\def\ergsec{{\rm erg\, s^{-1}}}

\slugcomment{Draft, 20100303}

\begin{document}

\title{
Radiative Transfer Modeling of Lyman Alpha Emitters. 
       I. Statistics of Spectra and Luminosity
}
\author{
Zheng Zheng\altaffilmark{1,2,3},
Renyue Cen\altaffilmark{4},
Hy Trac\altaffilmark{5},
and
Jordi Miralda-Escud\'e\altaffilmark{6,7}
}
\altaffiltext{1}{Yale Center For Astronomy and Astrophysics, Yale University, 
                 New Haven, CT 06520; zheng.zheng@yale.edu
}
\altaffiltext{2}{Institute for Advanced Study, Einstein Drive,
                 Princeton, NJ 08540
}
\altaffiltext{3}{John Bahcall Fellow}
\altaffiltext{4}{Department of Astrophysical Sciences, Princeton University,
                 Peyton Hall, Ivy Lane, Princeton, NJ 08544
}
\altaffiltext{5}{Harvard-Smithsonian Center for Astrophysics,
                 Cambridge, MA 02138
}
\altaffiltext{6}{Instituci\'o Catalana de Recerca i Estudis Avan\c cats,
                 Barcelona, Spain
}
\altaffiltext{7}{Institut de Ci\v cncies del Cosmos, Universitat de Barcelona, 
                 Barcelona, Spain
}

\begin{abstract}

We combine a cosmological reionization simulation with box size of 100$\hMpc$
on a side and a Monte Carlo \lya radiative transfer code to model 
Lyman Alpha Emitters (LAEs) at $z\sim$5.7. The model introduces \lya radiative
transfer as the single factor for transforming the intrinsic \lya emission
properties into the observed ones. Spatial diffusion of \lya photons from 
radiative transfer results in extended \lya emission and only the central part 
with high surface brightness can be observed. Because of radiative transfer, 
the appearance of LAEs depends on density and velocity structures in 
circumgalactic and intergalactic media as well as the viewing angle, which 
leads to a broad distribution of apparent (observed) \lya luminosity for a 
given intrinsic \lya luminosity. Radiative transfer also causes frequency 
diffusion of \lya photons. The resultant \lya line is asymmetric with a red 
tail. The peak of the \lya line shifts towards longer wavelength and the shift 
is anti-correlated with the apparent to intrinsic \lya luminosity ratio. The 
simple radiative transfer model provides a new framework for studying LAEs. 
It is able to explain an array of observed properties of $z\sim$5.7 LAEs in 
Ouchi 
et al. (2008), producing \lya spectra, morphology, and apparent \lya luminosity 
function (LF) similar to those seen in observation. The broad distribution of 
apparent \lya luminosity at fixed UV luminosity provides a natural explanation 
for the observed UV LF, especially the turnover towards the low luminosity end.
The model also reproduces the observed distribution of \lya equivalent width 
(EW) and explains the deficit of UV bright, high EW sources. Because of the 
broad distribution of the apparent to intrinsic \lya luminosity ratio, the 
model predicts effective duty cycles and \lya escape fractions for LAEs.

\end{abstract}

\section{Introduction}

More than four decades ago, \citet{Partridge67} proposed that prominent
\lya emission reprocessed from ionizing photons 
of young stars in galaxies can be used to detect high-redshift galaxies. 
The first successful detections of 
high-redshift \lya emitting galaxies, or \lya emitters (LAEs), were made 
$\sim$ 30 years later \citep[e.g.,][]{Hu96,Cowie98,Dey98,Hu98,Hu99}.
Recently, important advances have been made on the observational 
front to detect LAEs at $z\gtrsim 6$
\citep[e.g.,][]{Hu98,Hu02,Hu04,Hu05,Hu06,Rhoads03,Malhotra04,Horton04,Stern05,
Kashikawa06,Shimasaku06,Iye06,Cuby07,Ouchi07,Ouchi08,Stark07a,Nilsson07,
Willis08,Ota08}. 

LAEs can be efficiently detected through narrow-band imaging or with 
integral-field-units (IFU) spectroscopy. Owing to the high efficiency of 
target detection, LAEs naturally become objects for large surveys of 
high-redshift galaxies. 
Besides providing clues to the formation and evolution of galaxies at the time
when the universe was still young, LAEs are an important tracer of
the large-scale structure. The clustering of LAEs
may be used to constrain cosmological parameters. In particular, the 
large-volume surveys such as the Hobby-Eberly Telescope Dark Energy Experiment 
(HETDEX; \citealt{Hill08}) will enable the detection of the baryon 
acoustic oscillations (BAO) features \citep[e.g.,][]{Eisenstein05} in 
the LAE power spectrum. The BAO and the shape of the power spectrum can be 
used to measure the expansion history of the universe at early epochs
($z\sim 3$), which constrains the evolution 
of dark energy and the curvature of the universe.

LAEs are also a key probe of the high-redshift intergalactic medium (IGM),
especially across the reionization epoch. The use of LAEs to learn about 
reionization has been the subject of intense study 
\citep[e.g.,][]{Rees98,Miralda98,Haiman99,Santos04,Haiman05,Dijkstra07a,
Wyithe07}. 
Suitably devised statistics, 
including luminosity function (LF) and correlation functions of LAEs, 
can be used to constrain the neutral fraction of the IGM during reionization
\citep[]{Malhotra04,Haiman05,Kashikawa06,Furlanetto06,Dijkstra07b,McQuinn07,
Mesinger08,Iliev08,Dayal08,Dayal09}. 
By comparing the LFs of $z\sim 5.7$ and $z\sim 6.5$ LAEs, \citet{Malhotra04} 
conclude that reionization was largely complete at $z\sim 6.5$ (also see 
\citealt{Dijkstra07b}). \citet{McQuinn07} show that, with the angular 
correlation function of the 58 available $z\sim 6.6$ LAEs in the Subaru Deep 
Field \citep{Kashikawa06}, limits may be placed on the IGM neutral fraction, 
favoring a fully ionized universe at $z\sim 6.6$.


However, none of the previous work of LAEs mentioned above used reionization 
simulations with concurrent treatment of hydrodynamics plus radiative 
transfer of ionizing photons and \lya photons.
Hydrodynamic and radiative transfer simulations provide realistic neutral 
gas distributions, and \lya radiative transfer yields detailed
properties of the \lya 
emission. Realistic \lya radiative transfer calculations have
been applied to high-redshift LAEs in cosmological simulations 
\citep[e.g.,][]{Tasitsiomi06}. The application, however, is limited to  
a few individual sources, which do not form a sample for statistical study.

\citet{McQuinn07} and \citet{Iliev08} studied a sample of LAEs in reionization 
simulations with cosmological volume. However, the radiative transfer
of \lya photons is treated in a simplistic way in their study: the 
observed \lya spectrum is modeled as the 
intrinsic line profile modified by $\exp(-\tau_\nu)$, where $\tau_\nu$ is 
the optical depth at frequency $\nu$ along the line of sight. Although this
$\exp(-\tau_\nu)$ model can yield insights into the properties of the
observed \lya emission, such as the effect of IGM on the observability of LAEs, 
it is far from a complete description of the radiative transfer of \lya 
photons. First, during the propagation, \lya photons experience frequency 
diffusion, which is neglected by the simple $\exp(-\tau_\nu)$ model. The 
$\exp(-\tau_\nu)$ model removes \lya photons at a given frequency according
to the \lya optical depth, and no frequency change occurs for any \lya 
photon, therefore it does not yield correct \lya spectra. Second, the
simple $\exp(-\tau_\nu)$ model does not account for the spatial diffusion
of \lya photons either. LAEs in this model appear as point sources in \lya 
and there is no surface brightness information. Even if \lya photons start
from a point source, spatial diffusion due to radiative transfer would 
lead to an extended source. Observationally, LAEs indeed appear to be extended 
and they are defined by a surface brightness threshold in the narrow-band 
image \citep[e.g.,][]{Ouchi08}. Therefore, although the simple
$\exp(-\tau_\nu)$ model may provide useful insight, it likely falls short
for predicting the detailed properties of the observed \lya emission from LAEs.

To correctly understand high-redshift LAEs and use them for cosmological 
study, a full calculation of radiative transfer of \lya photons for a large 
sample of LAEs in cosmological reionization simulation is necessary, as will
be evident later.
In this work, we aim to perform detailed radiative transfer calculation of 
\lya photons \citep{Zheng02} from LAEs in a self-consistent fashion through
radiation-hydrodynamic reionization simulations \citep{Trac08}. For this paper,
we focus on studying statistical properties of $z\sim 5.7$ LAEs and show how 
the radiative transfer calculation aids our understanding of the observed 
properties of LAEs. The clustering properties of LAEs from this study will be 
presented in another paper (Paper II; Zheng et al. in prep.). The paper is 
organized as follows. In \S~2 we review the cosmological reionization 
simulation used in this work and in \S~3 we describe the \lya radiative 
transfer calculation. 
In \S~4, we study in details the \lya emission from an individual source 
chosen from the simulation box to gain a general view of the effect of \lya 
radiative transfer on the appearance of LAEs. Then, we present the statistical
properties of LAEs in \S~5, including their spectra and luminosity, from our 
modeling of an ensemble of sources in the simulation box. We compare our
modeling results with observations for $z\sim 5.7$ LAEs and discuss the 
implications in our understanding of LAEs. \S~7 is devoted to identifying 
important physical factors in shaping the observed \lya emission of LAEs.
We summarize and discuss the results in \S~8.

\section{Radiation Hydrodynamic Simulation of Cosmological Reionization}

In this work, we perform a \lya radiative transfer calculation to model LAEs.
The sources and physical properties of gas are taken from the outputs of
a cosmological reionization simulation.

The cosmological simulation \citep{Trac08} models cosmic reionization 
by using a 
hybrid approach to solve the coupled evolution of the dark matter, baryons, 
and radiation (\citealt{Trac04,Trac06,Trac07}). First, a high-resolution 
$N$-body simulation was run, with 3072$^3$ dark matter particles on a mesh of 
11,520$^3$ cells in a box of 100$\hMpc$ (comoving) on a side, and 
collapsed dark matter halos were identified on the fly. These halos are
the sites to form sources of ionizing photons. The high resolution and large 
box size of the simulation make it possible to resolve small scale structures 
and to reduce sample variance for source statistics. 

Hydrodynamics and radiative transfer of ionizing photons are simulated with 
moderate resolution (equal numbers, 1536$^3$, of dark matter particles, gas 
cells, and adaptive rays). 
Within the limits of available computational resources, the multi-grid 
approach adopted in the simulation maximizes the resolution of the individual 
numerical components (gas and radiation) in order
to model the corresponding physics adequately.
The initial conditions are the same as in the high-resolution $N$-body 
simulation, and the high-resolution simulation is used only to generate a 
catalog of halos at each redshift step and to obtain the list of sources of 
ionizing radiation. These sources for the ionizing photons are then used in 
the lower resolution simulation. The sources are assumed to be Population II
stars from starbursts \citep{Schaerer03}, and they are related to halos
according to the prescription for star formation and emitted radiation
in \citet{Trac07}.
For each gas cell, the incident 
radiation flux is used to solve the temperature and ionization structure 
of each cell. For more details about the simulation, see \citet{Trac07} and 
\citet{Trac08}. 

The simulation adopts a spatially flat 
$\Lambda$CDM cosmological 
model with Gaussian initial density fluctuations, and the cosmological 
parameters are consistent with the Wilkinson Microwave Anisotropy Probe (WMAP) 
5-year data \citep{Dunkley09}: $\Omega_m=0.28$, $\Omega_\Lambda=0.72$,
$\Omega_b=0.046$, $h=0.70$, $n_s=0.96$, and $\sigma_8=0.82$.


\section{\lya Radiative Transfer Calculation}
\label{sec:RT}

The outputs of the above simulation form the basis for computing the
radiative transfer of \lya photons and studying LAEs. The radiative
transfer of the resonance \lya line has been a subject of intense
study \citep[e.g.,][]{Hummer62,Auer68,Avery68,Adams72,Harrington73,
Harrington74,Neufeld90,Neufeld91,Loeb99,Ahn00,Ahn01,Ahn02,Zheng02,
Dijkstra06,Hansen06,Tasitsiomi06,Verhamme06,Laursen09,Pierleoni09}.
Owing to the complex nature of the geometry and gas distribution
in the cosmological realization we study, the
Monte Carlo method of solving the \lya radiative transfer becomes the
natural choice. We use the Monte Carlo code developed in \citet{Zheng02}, 
modified to use the simulation output, to solve the \lya radiative transfer 
in this study. This code has also been applied to study the fluorescent
\lya emission from the IGM in a hydrodynamic simulation \citep{Kollmeier10}.

The code works as follows.
\begin{itemize}
\item[(1)] 
    For each \lya photon, its initial position is drawn according to the 
    emissivity distribution in the box (a superposition of point sources 
    in the case presented in this paper, see below). The initial frequency 
    of the photon follows the Gaussian distribution determined by the 
    halo virial temperature (see below) in the rest frame of the fluid at the 
    photon's position and its direction is randomly distributed.
\item[(2)] 
    An optical depth is then drawn from an exponential distribution. The 
    spatial location along the chosen direction corresponding to this optical 
    depth is determined from the distributions of neutral hydrogen density, 
    fluid velocity, and temperature along this direction.
\item[(3)] 
    At this location, the \lya photon encounters a scattering. The frequency 
    and direction after the scattering are computed in the rest frame of 
    the hydrogen atom and then transferred back to the laboratory frame. 
\item[(4)]
    With the new frequency and direction, (2)--(4) are repeated until the
    photon escapes from the system (see below).
\end{itemize}
\lya photons are collected onto a three-dimensional (3D) array, which 
records the \lya spectra at each projected spatial location.
At each scattering, as well as at the initialization, the possibility that
the \lya photon escapes along the observational direction is computed and
added into the array. In the end, the output array of the Monte Carlo \lya 
radiative transfer code forms an IFU-like data cube.  \lya spectra 
(either 1D or 2D) can be extracted from this data cube and \lya images can 
be obtained by collapsing the data cube along the spectral direction.
For more details of the radiative transfer calculation, we refer the readers
to \citet{Zheng02}.


In this paper, we focus on the simulation output at $z=5.7$. 
The reionization is complete by that time in the simulation and the
neutral hydrogen fraction is about $9\times 10^{-5}$ in the IGM.
The neutral hydrogen density, temperature, and peculiar velocity fields
in the simulation box are stored in a 768$^3$ grid, which 
feeds to the \lya radiative transfer code. The Hubble flow is added to the 
velocity field. LAEs are modeled to reside in dark matter halos. The 
positions and velocities of LAEs are from the halo list. 
To reduce source blending in the \lya image and spectra, \lya photons
are collected with a spatial resolution finer than the above grid, with
each grid resolved by 8$^2$=64 pixels. The size of each pixel corresponds to 
16.3$\hkpc$ (comoving) or 0.58\arcsec.
The resolution of the 768$^3$ grid for gas properties used in the \lya 
radiative transfer calculation is a factor of two lower than in the 
hydrodynamical simulation, as a result of computational efficiency 
consideration. The slight smoothing 
of gas fields may cause a smoothing effect in \lya surface brightness profile.
However, since we use a finer grid to collect \lya photons and \lya sources 
are initially point sources (see below), the smoothing in \lya image is 
expected to be much weaker than the smoothing in gas properties.

The whole simulation box is divided
into three layers along the line of sight, 
with the volume of each layer being 
100$\times$100$\times$33.33 $\VhMpc$. The depth of each layer, 33.33$\hMpc$,
is close to the width of the narrow-band filter used to search for 
$z\sim$5.7 LAEs in the 1 deg$^2$ field of the Subaru/{\it XMM-Newton}
Deep Survey (SXDS), and the area is almost identical to that of the survey as 
well \citep{Ouchi08}. Therefore, we have three SXDS-like volumes at $z\sim$5.7.
For each layer, the output array on which the \lya photons are collected
has a total range of 24\AA (rest frame) along the spectral
direction, a width large enough to cover the Hubble expansion plus the
peculiar velocities in the $33.33 \hMpc$ width of each layer.
The spectral resolution is set to be 0.1\AA (rest frame), corresponding
to 25$\kms$. As a whole, the \lya radiative transfer results for each layer 
are saved in an array of dimension 6144$\times$6144$\times$240. 

\begin{figure*}
\plotone{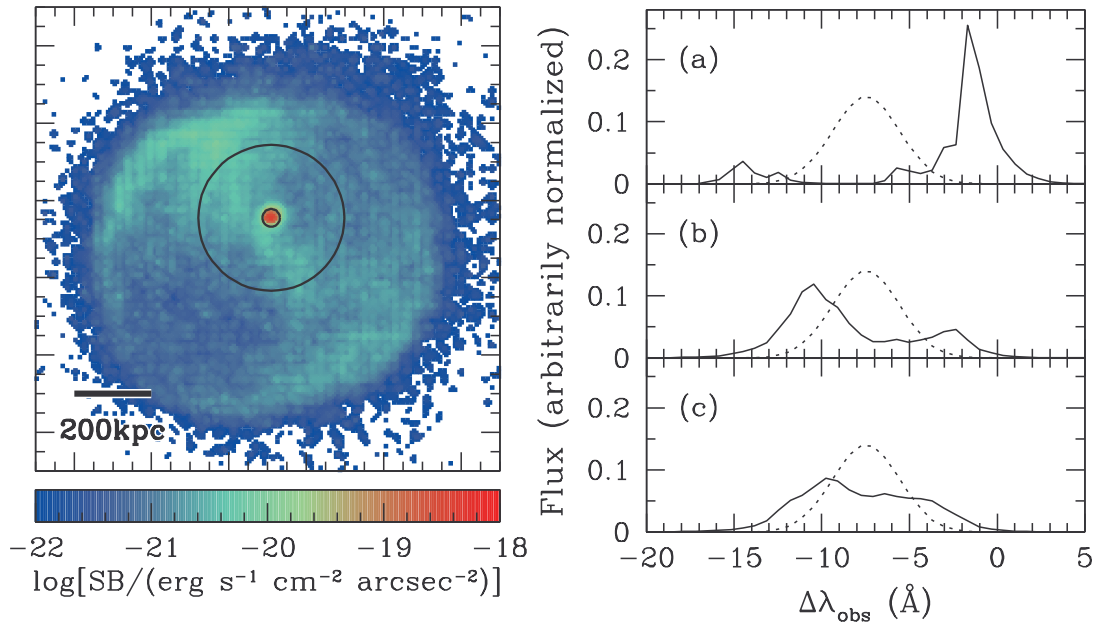}
\caption[]{
\label{fig:imgspec}
\lya image and spectra of a single $z\sim 5.7$ LAE, randomly chosen in the 
simulation box. {\it Left:} The \lya surface brightness distribution after 
the radiative transfer. The physical size of the region is indicated by the 
scale bar and the separation of the two adjacent ticks on the axes correspond 
to 50kpc. Two circles delineate the apertures for the spectra in the right 
panels.  The virial radius of the host halo ($\sim 10^{11}\hMsun$) is
about 26kpc, slightly larger than that of the inner circle.
{\it Right:} \lya spectra (solid curves) at different radii to the 
source center. Panel ($a$) shows the spectrum within the smaller circular 
aperture shown in the image, and Panel ($b$) for photons in between 
the smaller and the larger circles and Panel ($c$) for photons outside of
the larger circle. The dotted curve in each panel shows the intrinsic \lya 
line profile, which would be observed if there were no scatterings. The 
spectra are arbitrarily normalized, since we concentrate on the profiles.
The wavelength is shown as the difference to $(1+z)\lambda_0$, where $z=5.7$
and $\lambda_0$=1216\AA is the rest-frame wavelength of \lya.
}
\end{figure*}

We perform the \lya scattering calculation for all the halos above $5\times
10^9\hMsun$. Assuming $\sim$2/3 of ionizing photons are converted to \lya
photons (case-B recombination, \citealt{Osterbrock89}) and a \citet{Salpeter55} 
initial mass function (IMF), the intrinsic \lya luminosity $\Lint$ is related 
to the star formation rate (SFR) as \citep{Furlanetto05}
\begin{equation}
\label{eqn:Lint}
\Lint=10^{42} [{\rm SFR}/(\Msun {\rm yr}^{-1})]\, \ergsec.
\end{equation}
In the simulation, the resultant SFR under the adopted star formation 
prescription \citep{Trac07}
is found to be tightly correlated with halo mass, 
\begin{equation}
\label{eqn:sfr}
{\rm SFR}=0.68 [M_h/(10^{10} h^{-1}M_\odot)]\, \Msun {\rm yr}^{-1}.
\end{equation}
So the intrinsic \lya luminosity and halo mass are almost interchangeable
in our model and in our descriptions of the results. 
The relation in equation~(\ref{eqn:sfr}) holds at $z\sim$5.7, and there is 
a redshift dependence (see \citealt{Trac07}). Since the ultra-violet
(UV) luminosity is also proportional to SFR, the halo mass to UV (or intrinsic 
\lya) light ratio is approximately constant in our model.

For each halo, \lya photons are launched at the halo center. The point source
assumption is reasonable. 
\lya emission originates from reprocessed ionizing photons
of massive stars \citep{Partridge67}. The ionizing photons ionize the 
neutral hydrogen atoms in the interstellar medium (ISM) and the case-B 
recombination has a probability of $\sim$2/3 of ending up as \lya photons
\citep{Osterbrock89}. We aim to solve the radiative transfer in the 
circumgalactic and intergalactic environments, and the initial \lya photons 
launched in our model correspond to photons just escaping from the ISM
whose spatial distribution closely follows the UV light of galaxies.
From HST/ACS observations of $z=5.7$ LAEs, 
\citet{Taniguchi09} find that in the broad-band (rest-frame UV) images,
LAEs are compact sources with sizes of less than 1 kpc,
smaller than the pixel size in our modeling.
Therefore, our assumption of a point source for the initial \lya emission
is justified.

The initial frequency of the \lya photons in the rest frame of the 
halos is assumed to follow a Gaussian 
distribution with the width corresponding to the virial temperature of the
halo, $T_{\rm vir}=GM_h\mu m_H/(3kR_{\rm vir})$ \citep{Trac07}, where $\mu\sim
0.59$ is the mean molecular weight. 
The rms line width is 
$\sigma_{\rm init}=31.9[M_h/(10^{10} h^{-1}M_\odot)]^{1/3} \kms$.
This width is about half of the circular 
velocity at the virial radius. This is a rather conservative assumption and we
will test and discuss the effect of increasing the initial width on our 
results. The total number of \lya photons drawn for each halo is 
$N_\gamma=$max\{SFR/($\Msun {\rm yr}^{-1}$),$10^3$\}, and these photons are
given a weighting factor $w$ to convert the photon number to the \lya 
luminosity of the halo, $\Lint=wN_\gamma$.

To fully account for the effect of the IGM on \lya 
radiative transfer, we impose the periodic boundary conditions of the 
simulation in our \lya radiative transfer calculation. For each \lya photon, 
we stop the scattering calculation when it reaches a distance of half of
the box size ($L=100\hMpc$) from the initial position in any of the three 
principle directions. At this distance, the Hubble expansion leads to a
fractional shift in \lya wavelength of the order of 
$\Delta\lambda/\lambda=0.5H(z)L/(1+z)/c\sim 1.6\times 10^{-2}$, corresponding
to a velocity of $\sim$5000$\kms$. This is much larger than typical values of
peculiar velocity of halos and the shift caused by frequency diffusion, and
is more than sufficient to ensure that the photon will no longer interact at
the \lya line. In fact, most of the time the photon is last scattered at a 
distance a few comoving Mpc away from the source (e.g., see 
Fig.~\ref{fig:imgspec}).

\section{Detailed Study of an Individual LAE}

\begin{figure*}
\plotone{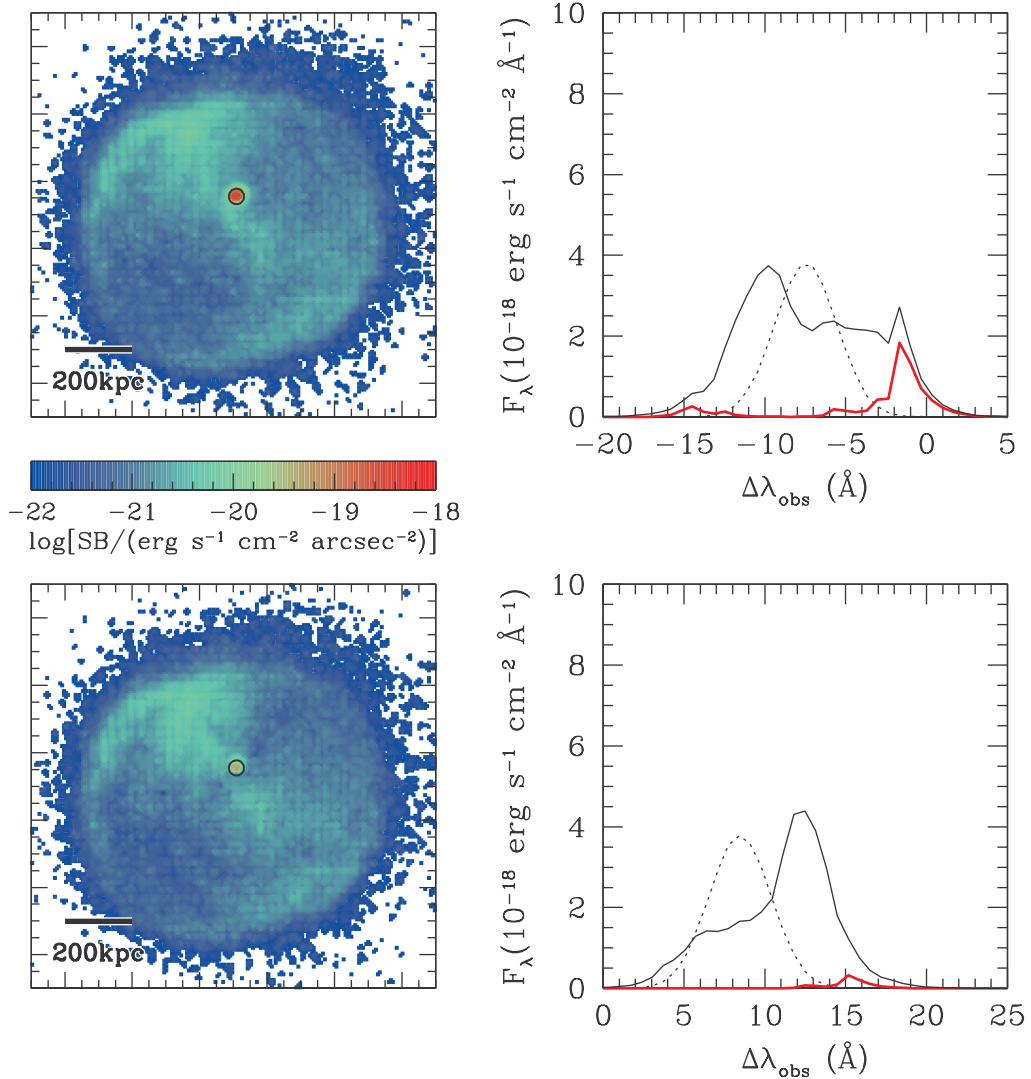}
\caption[]{
\label{fig:diffview}
\lya images ({\it left}) and spectra ({\it right}) of a single $z\sim 5.7$ 
LAE observed along two opposite directions. Mirroring reflection has been 
applied so that the two images have the same orientation.
In the image panels, the physical size of the region is indicated by the
scale bar and the separation of the two adjacent ticks on the axes correspond 
to 50kpc.
In each of the right panel, the dotted curve shows the intrinsic \lya
line profile, which would be observed if there were no scatterings.
The solid black curve shows the spectrum of all scattered photons
and the solid red curve is the spectrum near the center (within the circular
aperture shown in the image panel). All quantities are in the observer's frame.
The wavelength is shown as the difference to $(1+z)\lambda_0$, where $z=5.7$
and $\lambda_0$=1216\AA is the rest-frame wavelength of \lya. The relative 
shift of wavelength ranges in the top and bottom spectra panels reflects the
change in the viewing direction, which leads to the change in the source 
position with respect to the box center and that in the velocity direction 
with respect to the observer.
}
\end{figure*}

\begin{figure*}
\plotone{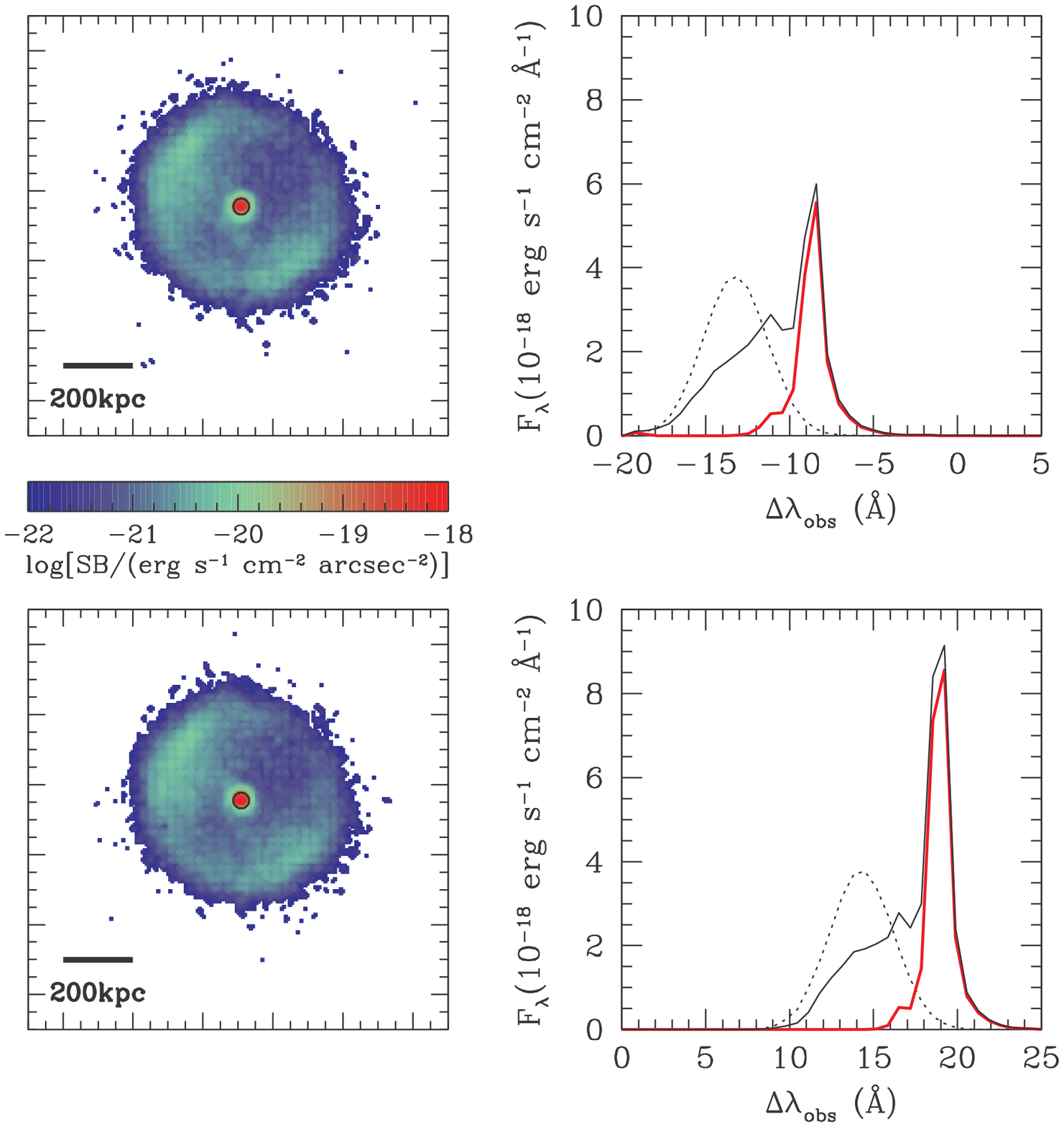}
\caption[]{
\label{fig:diffview_zerovpec}
Same as Fig.~\ref{fig:diffview} but with the peculiar velocity field turned off.
}
\end{figure*}

Before presenting \lya radiative transfer and statistical results 
for all the LAE sources in the whole simulation box, we first examine  
in detail the radiative transfer results for an individual 
source to aid our understanding of the general effects of \lya 
scattering. 

We randomly chose a halo in the simulation box, which has a mass of
$\sim 10^{11}\hMsun$.
The size of the virialized halo is about 26kpc, slightly larger than
the inner circle in the left panel of Figure~\ref{fig:imgspec}.
As mentioned in \S~\ref{sec:RT}, \lya photons are 
assumed to initially start from a point source, located at the halo center. 
The left panel of Figure~\ref{fig:imgspec} shows the \lya surface brightness 
distribution after \lya scattering from this source. Because of the radiative 
transfer, the initial point source becomes an extended source and a roughly 
spherical halo of scattered \lya photons emerges. This scattered \lya halo is 
similar to the one around a point source before reionization described in 
\citet{Loeb99}. While \citet{Loeb99} assume a uniform, zero temperature IGM 
undergoing Hubble expansion, we use a realistic distribution of gas density, 
temperature, and velocity around a star-forming halo. This causes deviations 
from spherical symmetry in the surface brightness profile. The scattered \lya 
surface brightness drops as the radius increases. 
The blue-green extended \lya emission seen in Figure~\ref{fig:imgspec} is out 
in the IGM, corresponding to the scattered photons as they travel to the 
region where the Hubble expansion compensates for their blueshift acquired by
the scatterings in the infall region of the halo. The sharp edge around a 
radius of $\sim$0.5Mpc reflects the frequency 
of the ``bluest'' photons coming out of the central (infall) region before 
encountering the IGM. The bluer the photons are, the farther they can travel 
in the IGM before redshifted to the line center and significantly scattered.
In practice, we can only 
observe the very inner part of the extended \lya radiation, where the surface 
brightness is high. The extended \lya halo would merge with those from 
neighboring sources, forming a \lya background. We will describe how we 
identify sources in the simulation in \S~\ref{sec:identification}.

In the right panels of Figure~\ref{fig:imgspec}, we show \lya spectra at 
different radii of the source. The dotted curve in each panel is the intrinsic
\lya line profile, assumed to be Gaussian with the width determined by the
virial temperature of the dark matter halo hosting the source. It would be the
observed spectrum if \lya photons streamed out of the source without any 
scatterings.
Note that the wavelength shown in the plot is the difference from 
$(1+z)\lambda_0$, where $z=5.7$ and $\lambda_0$=1216\AA is
the rest-frame 
wavelength of \lya. The offset of the peak of the initial line profile 
from zero is caused by a combination of the Hubble velocity with respect
to the center of the simulation box and the peculiar velocity of the
source.
On average, \lya photons are first scattered by neutral hydrogen atoms in the 
infall region around the LAE host halo (see \S~\ref{sec:factors} and 
Fig.~\ref{fig:halo_profile}). 
The inner infall region has an inverted
Hubble-like contraction velocity distribution and \lya photons escaping at the
radius of maximum infall velocity have their frequency most likely shifted to 
the blue side of the line center \citep{Zheng02}. Then \lya photons experience 
scatterings in the region with decreasing infall velocity (outer infall region)
and finally hit the Hubble flow. The subsequent scatterings on average shift 
the frequency of \lya photons redward. 

Most of the observed \lya photons from regions with high surface brightness 
near the center of the source shift to the red side (Fig.~\ref{fig:imgspec}$a$)
with respect to the intrinsic distribution. These photons
are most likely to have had forward scatterings along the line of sight. The 
outward increasing gas velocity (from the outer infall region and the Hubble 
flow) makes it easier for photons that have redward frequency shifts to escape.
At larger lateral radii, (Fig.~\ref{fig:imgspec}$b$ and 
Fig.~\ref{fig:imgspec}$c$), we have more contributions from 
photons that travel perpendicular to the line of sight but are scattered to
the line-of-sight direction. The frequency after scattering would be near the
incoming frequency seen in the rest-frame of the atom at the scattering radius.
Since most of the scatterings would happen around the radius where the Hubble 
expansion velocity redshifts the \lya photons to the line-center frequency and
the line-of-sight neutral column density become smaller at larger radii, the
observed photons at large projected radii would appear bluer than those 
observed near the center, as seen in the spectra at large radii.

Only the very central part of the extended \lya radiation can be 
practically observed as the LAE source. The transmitted \lya flux depends on the
gas distribution and kinematics in the halo vicinity, and the viewing angle. 
We perform a few tests to see the influence of the gas properties in the 
observed flux of the central part.
Figure~\ref{fig:diffview} compares \lya images and spectra of the above source
observed in two opposite directions. Mirroring reflection has been applied 
to one image so that the two images have the same orientation. 
While the two images have similar spatial extent, there is a large difference (about a factor
of 7) in the flux inside the central aperture. The difference can be clearly
seen from the spectra. The red curve in each of the right panels is the
spectrum extracted from the circular aperture near the source center. 

Evidently, the differences in the intervening gas distribution,  
namely the neutral hydrogen density and peculiar velocity distributions,
have a dramatic effect on the observed \lya flux.
To test this we perform a scattering calculation
with the peculiar velocities of the source and neutral gas set to zero,
while keeping the Hubble expansion. 
Figure~\ref{fig:diffview_zerovpec}
compares the resultant images and spectra observed in the two opposite 
directions. Any remaining differences between the results of different viewing
directions should be caused only by the anisotropic density or temperature 
distribution around the source. 
We see that, with the peculiar velocity field turned off,
the difference in the fluxes from the central aperture 
between the two lines of sight becomes much smaller, 
a factor of $\sim 1.5$ (versus $\sim 7$ in the case with peculiar velocity). 
The spectra from the central aperture also look more similar to each other. 

With the peculiar velocity turned off, the surface brightness profile
of scattered \lya photons appears to be more concentrated than that in 
Figure~\ref{fig:diffview}. This is mainly a consequence of the disappearance 
of the infall region around the source by artificially setting the peculiar 
velocity to zero. If the peculiar velocity is not switched off, \lya photons 
climbing out of the infall region (before reaching the IGM dominated by the 
Hubble expansion) would on average have shifted blueward \citep{Zheng02} 
with respect to the line center. Compared to the case without the 
blueward shift (e.g., the initial Gaussian profile when the peculiar velocity 
is turned off), these bluer photons would travel a larger distance in the IGM
before redshifting back to the line center and experiencing strong scatterings.
Therefore, we see a more extended \lya emission in the case with the peculiar
velocity. 

The above tests show that peculiar velocity plays an important role in the
scattered \lya brightness profile and the transmitted flux near the source 
center. In \S~\ref{sec:factors}, in additional to the peculiar velocity, we 
identify other factors in affecting the \lya transmission and statistically 
study their role in the observability of LAEs.

\begin{figure*}
\epsscale{1.2}
\plotone{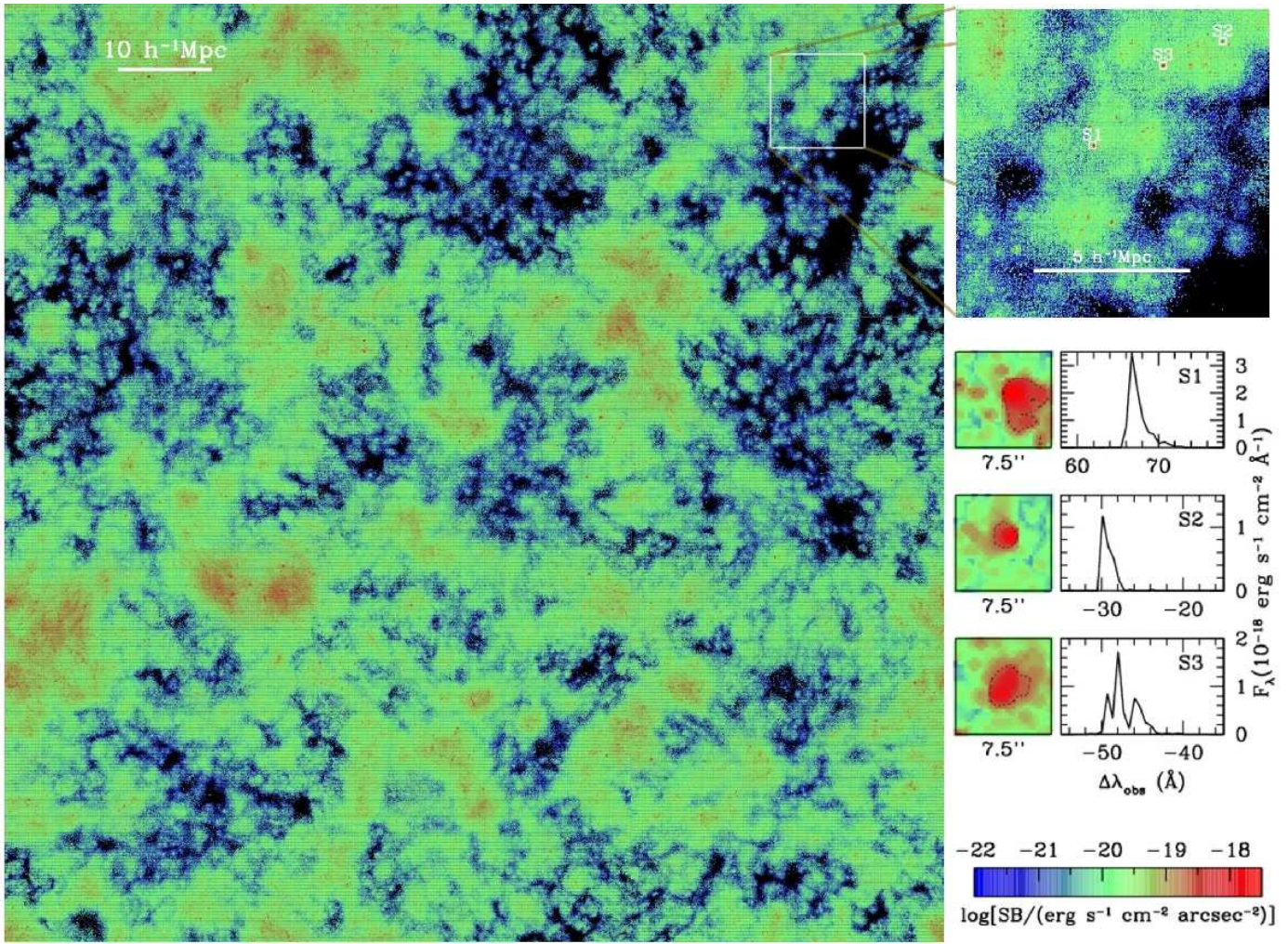}
\epsscale{1.0}
\caption[]{
\label{fig:mosaic}
Scattered \lya emission from LAEs. The \lya surface brightness distribution
is shown for sources in one third of the simulation box, with 100$\hMpc$
on a side and a thickness of 33.33$\hMpc$. The area matches that of the SXDS
and the depth corresponds to the width of the narrow-band filter for 
$z\sim 5.7$ LAEs \citep{Ouchi08}. The morphology and spectra of three sources 
in the zoom-in region are shown.
}
\end{figure*}

\begin{figure*}
\plotone{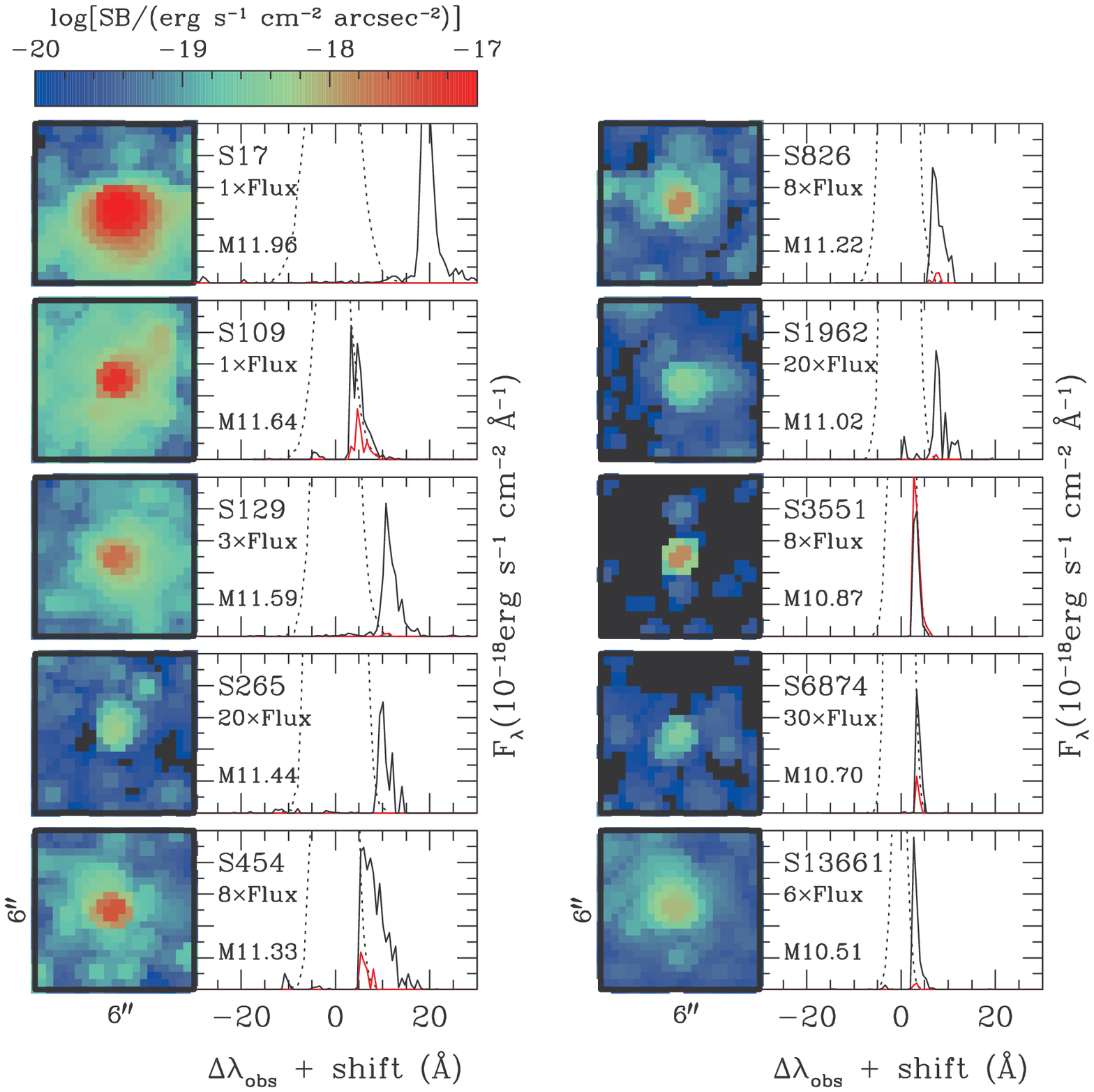}
\caption[]{
\label{fig:imgspec_sel}
Images and spectra of a few $z\sim 5.7$ LAE sources in our model. 
In each of the panels of spectra, the dotted curve is the intrinsic line
profile, which would be observed if \lya photons escaped without scattering.
The black solid curve is the spectrum in our model with the full \lya 
radiative transfer. For comparison, the solid red curve is the spectrum 
with a simple treatment of radiative transfer, which modifies the intrinsic
profile by multiplying $\exp(-\tau_\nu)$ with $\tau_\nu$ the line-of-sight
optical depth. For an easy comparison across panels, in each panel, all 
three spectra have the same constant horizontal shift so that the line center 
of the intrinsic profile is at zero. In each spectra panel, the label
starting with ``M'' denotes $\log(M_h)$, where the host halo mass $M_h$ is in 
units of $\hMsun$.
See text for details.
}
\end{figure*}

The results from the individual LAE source show that, as a consequence of 
radiative transfer, \lya photons experience both spatial diffusion and
frequency diffusion. An intrinsic point source of \lya emission appears 
extended and the \lya spectra differ substantially from the intrinsic 
Gaussian profile. The spectra from the central aperture, which is the part 
that is most observable, do not have a simple and clear relation to the 
initial line profile, owing to the frequency shift caused by scatterings. In 
some previous work (e.g., \citealt{McQuinn07,Iliev08}), the observed \lya 
spectrum is modeled as the intrinsic Gaussian profile modified by 
$\exp(-\tau_\nu)$ with $\tau_\nu$ being the optical depth at frequency $\nu$ 
along the line of sight. Such a simple model does not account for the 
frequency and spatial diffusion of \lya photons caused by radiative transfer. 
The resultant line profile in this simple model looks like a truncated 
Gaussian profile, with only the red tail transmitted. Our detailed \lya 
radiative transfer, on the other hand, shows that the observed \lya line 
profile near the source center is more complicated and the redward frequency 
shift is more than that in the simple treatment. The simple radiative transfer 
model may yield trends in some results that are qualitatively in accord with 
detailed transfer calculations. For example, \citet{Iliev08} also find that 
peculiar velocity is important in determining the observability of LAEs.
However, as we show in \S~\ref{sec:statistics}, the lack of frequency and 
spatial diffusion in the simple model means that it cannot capture the full 
picture of \lya emission from LAEs for detailed prediction and understanding of 
observed \lya features.

\section{Statistical Properties of \lya Spectra and Luminosity of LAEs}
\label{sec:statistics}

We perform \lya scattering calculation for all the sources residing in halos 
above $5\times 10^{9}\hMsun$ in the whole $(100\hMpc)^3$ box. In this section,
we describe how we identify LAEs from the post-scattering outputs and study
their statistical properties.

\subsection{Source Identification}
\label{sec:identification}

Figure~\ref{fig:mosaic} shows the \lya image for sources in one-third of the 
simulation box. It has an area of 100$\hMpc\times$100$\hMpc$ and a thickness
of 33.33$\hMpc$. The slice matches the sky coverage (1 deg$^2$) of the SXDS
and the depth corresponds to the width of the narrow-band filter 
($\Delta\lambda$=120\AA; \citealt{Ouchi08}) for $z=5.7$ LAEs. Therefore, 
the image can be regarded as an idealized, continuum-subtracted narrow band 
image of the $z=5.7$ LAEs for SXDS-like sky coverage and depth. From the whole
simulation box, we have three realizations of such a survey. The periodic
boundary condition of the simulation is imposed in our modeling, which can 
be clearly seen in Figure~\ref{fig:mosaic}. 

Because of \lya radiative transfer, LAEs are no longer point sources in our 
model. We need to find a way to define the sources in order to study their
statistical properties. Our identification of sources is motivated by the
procedure used in detecting LAEs in real observations. For $z=5.7$ LAEs in 
the SXDS, a threshold surface brightness of $2.64\times 10^{-18} {\rm erg\, 
s^{-1}\, cm^{-2}\, arcsec^{-2}}$ in the narrow-band image (including continuum) 
is adopted for detecting them (M. Ouchi, private communication). LAEs are 
identified by grouping pixels above this threshold. The observed rest-frame 
\lya equivalent width distribution of $z\sim 5.7$ LAEs peaks around 60\AA 
and is skewed to large values (Fig.23 in \citealt{Ouchi08}). Since the 
rest-frame width of the narrow-band filter is 120\AA/$(1+z)$, the continuum 
contribution to the surface brightness are likely to be less than 30\%. 
Our model does not include the continuum component. In principle, we could 
model the continuum based on the star formation history in the simulation, 
but the correction is small and it is not the main uncertainty of our model 
(as shown later in this paper). Therefore, we simply make a correction of 1/3
to remove the continuum contribution to the threshold surface brightness and
adopt $1.80\times 10^{-18} {\rm erg\, s^{-1}\, cm^{-2}\, arcsec^{-2}}$ as the
continuum-subtracted surface brightness for detecting SXDS LAEs.

\begin{figure*}
\plottwo{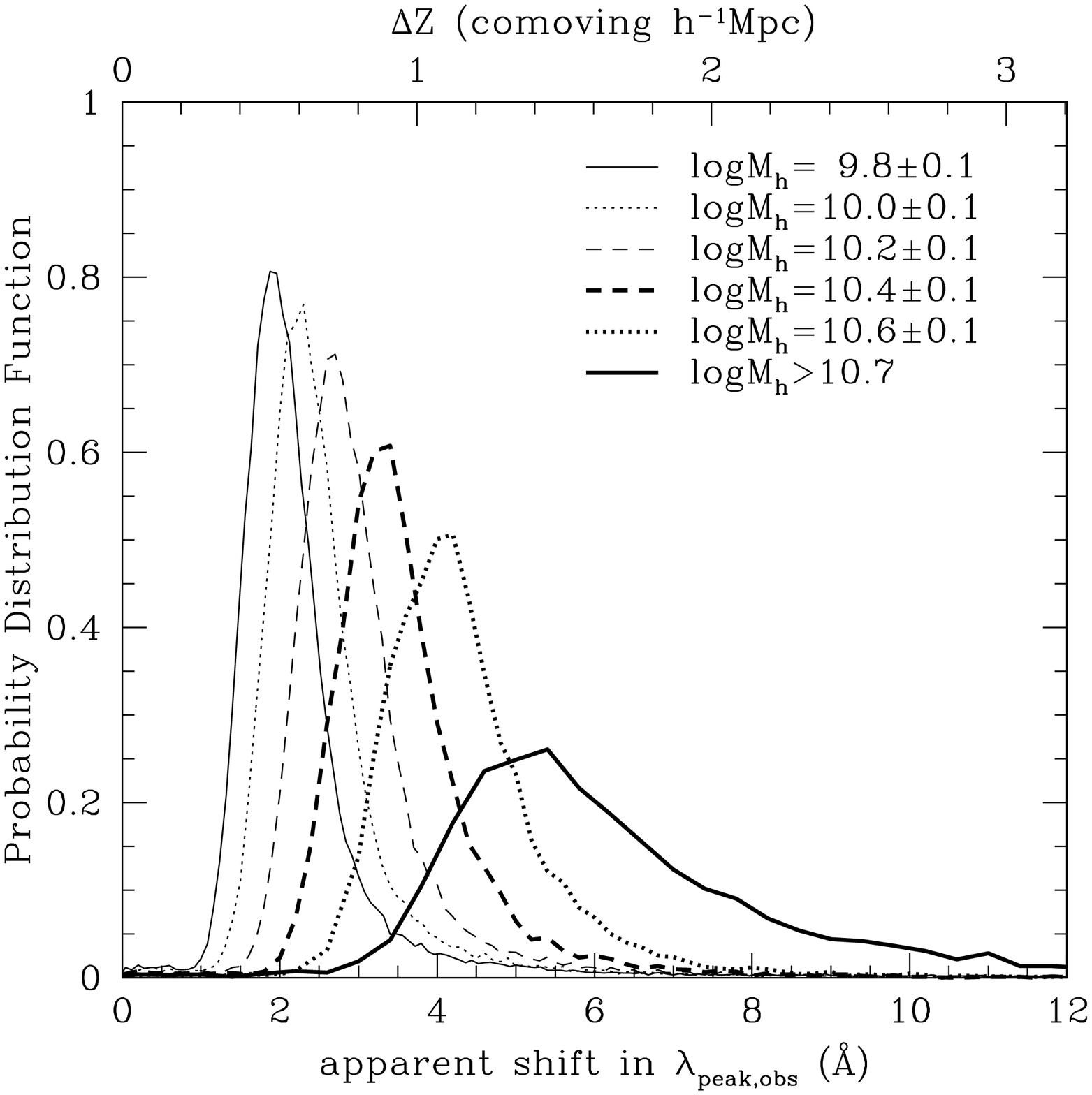}{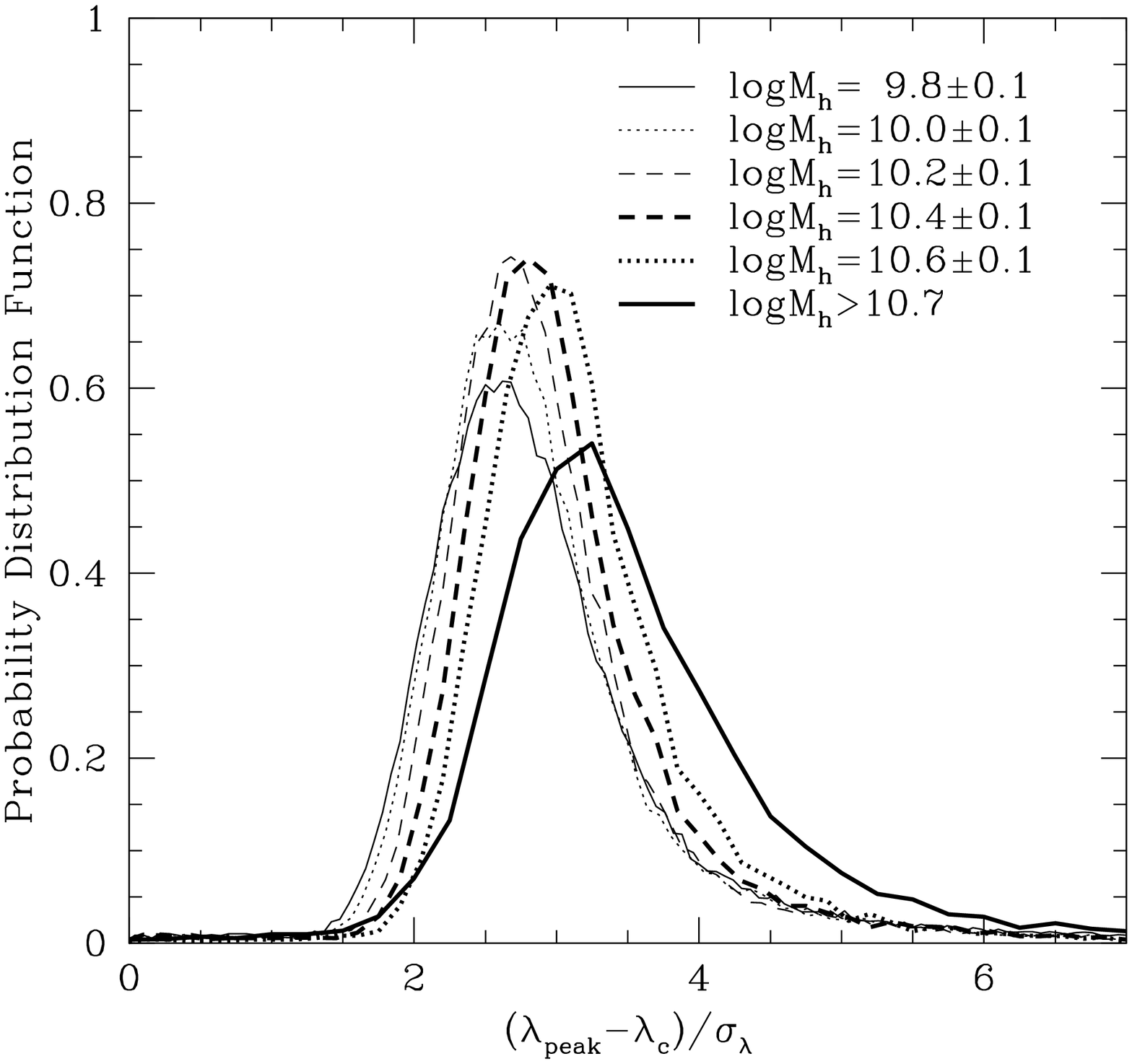}
\caption[]{
\label{fig:peakshift}
The apparent shift in the LAE \lya line peak with respect to the intrinsic 
one. Because of radiative transfer, the wavelength $\lambda_{\rm peak}$ at 
the peak flux of the observed \lya line is not $\lambda_c$ that corresponds 
to the center of the intrinsic (Gaussian) line profile. The apparent shift is 
defined as the wavelength difference. 
The top axis of the left panel marks the shift in 
comoving unit, which is of the order of $\hMpc$. 
{\it Left:} The distribution of the apparent shift of the peak wavelength 
as a function of LAE host halo mass (in units of $\hMsun$). 
The apparent wavelength shift can
translate to an apparent shift in the redshift-space source position $\Delta Z$
along the line of sight, which is labeled in the top axis.
{\it Right:} Similar to the left panel, but the peak wavelength shift is
in units of the intrinsic line width $\sigma_\lambda$, which is set by
the virial temperature of the host halo in our model.
}
\end{figure*}

\begin{figure}
\plotone{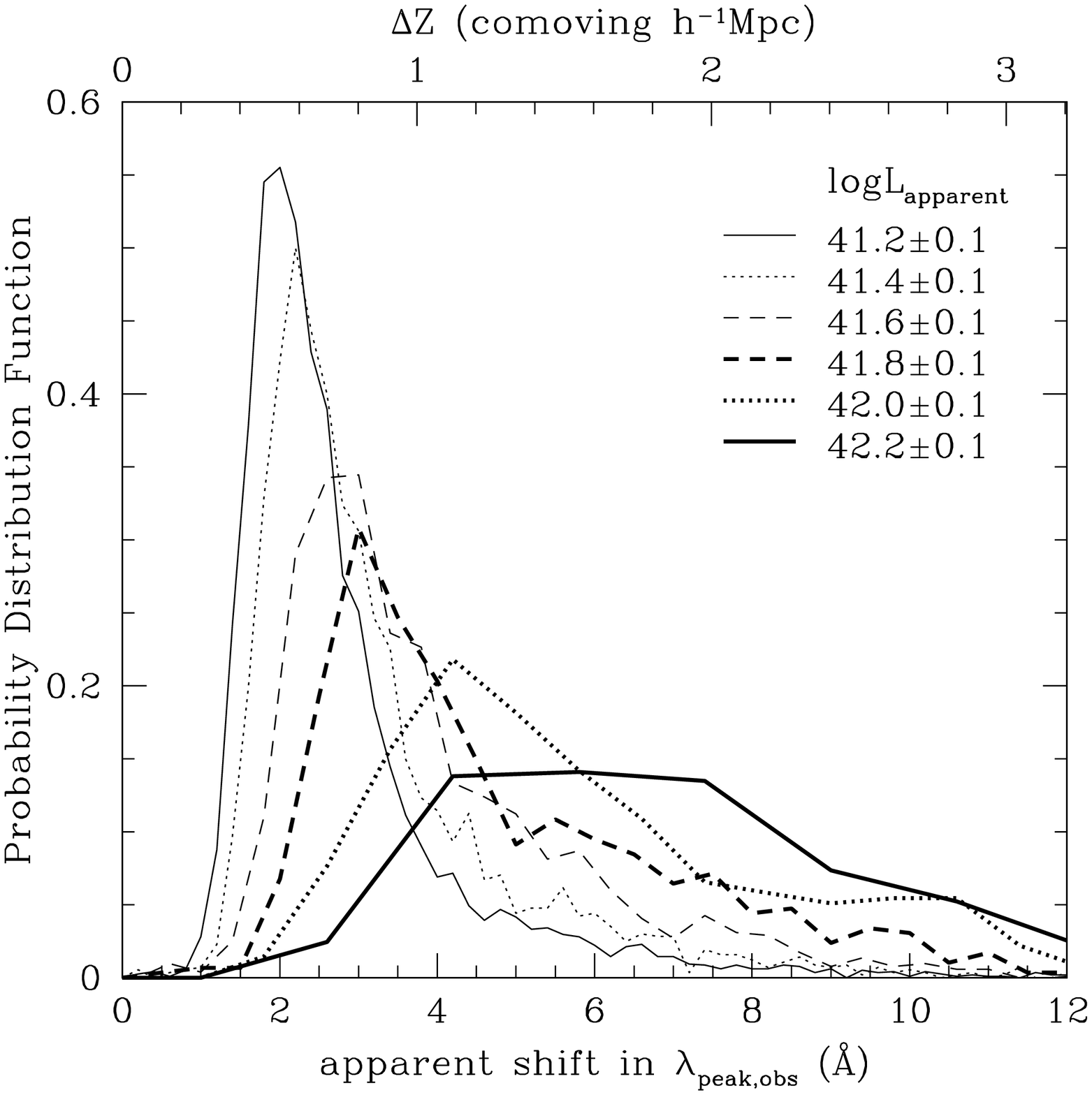}
\caption[]{
\label{fig:peakshift_obs}
Similar to the left panel of Fig.~\ref{fig:peakshift}, but for the apparent 
shift in the LAE \lya line peak as a function of the observed (apparent) 
\lya luminosity (in units of $\ergsec$).
}
\end{figure}

As shown in \S~\ref{sec:observation}, there is significant uncertainty in 
modeling \lya luminosity. To be conservative, we set a lower threshold surface 
brightness, $1.8\times 10^{-19} {\rm erg\, s^{-1}\, cm^{-2}\, arcsec^{-2}}$, 
for identifying LAEs
in our model. To have a full picture of the observability of LAEs (e.g., 
comparison between intrinsic and observed \lya emission), we also 
include the halo position information in the source identification. For 
a source with the projected position known from the halo catalog, we first
find the corresponding pixel in the \lya image. Starting from this root pixel, 
we then link the surrounding pixels with surface brightness above the 
threshold through a Friends-of-Friends (FoF) algorithm with the linking 
length equal to the size of the pixel. The directions to link the pixels 
are only horizontal and vertical in the image.
An LAE source is then defined by all the
linked pixels, and the spectra at each pixel's position are added together
to form the spectra of the source. In the case that the root pixel and its 
surrounding pixels all have surface brightnesses lower than the threshold, 
the flux and spectra from this root pixel are adopted for the source. As with 
other applications of the FoF algorithm, there are chances that two individual 
sources are bridged together. In this case, we also assign the flux and 
spectra of the root pixel to the intrinsically fainter source. Such cases
are rare and the correction does not affect any of our statistical study.

Figure~\ref{fig:imgspec_sel} shows \lya images and spectra of a few 
$z\sim 5.7$ LAEs from our model. 
The host halo masses of these sources are above $3\times 10^{10}\hMsun$. 
The corresponding intrinsic \lya luminosities are above 
$2\times 10^{42}\ergsec$ (equations~[\ref{eqn:Lint}] and [\ref{eqn:sfr}]),
roughly in the luminosity range probed by current LAE surveys like SXDS.
From top to bottom panels and left to 
right panels, they are arranged in order of decreasing intrinsic \lya 
luminosity (halo mass). Most of the sources appear to be roughly round with 
faint 
substructures around them, which are a combination of reflected \lya emission 
by clumps/filaments 
of neutral gas or \lya emission from fainter sources. The sizes and 
morphologies of the LAEs in our model are remarkably similar to those 
in the narrow-band images of $z\sim 5.7$ LAEs in SXDS (e.g., Fig.5 of 
\citealt{Ouchi08}; narrow-band images in Fig.3 of \citealt{Taniguchi09}).

Solid curves in the spectrum panels in Figure~\ref{fig:imgspec_sel} are the 
corresponding spectra for the shown LAEs. 
In each spectrum, the \lya line 
is clearly asymmetric, skewed towards the red. The line profiles resemble
the observed ones for the SXDS $z\sim 5.7$ LAEs (Fig.5 of \citealt{Ouchi08}).
However, the observed \lya lines appear to be much broader with less 
sharp blue edges. The difference can be simply attributed to the
spectral resolution: in the observer's frame, the observation typically has a 
resolution 8--15\AA \citep{Ouchi08}, while the resolution for our modeled 
spectra is 0.67\AA. 

The \lya lines with the full radiative transfer
show a clear distinction from the lines with a simple treatment of the
radiative transfer, namely the $\exp(-\tau_\nu)$ model.
For each source, the red curve is the \lya spectrum from the 
$\exp(-\tau_\nu)$ model, which is essentially the intrinsic Gaussian profile 
truncated below a certain wavelength. Although it displays a similar asymmetry,
the flux is usually significantly lower than that with the full radiative 
transfer. Importantly, the \lya line from the full radiative transfer model 
has a larger redward shift than that in the simple model, an effect that can 
only be properly modeled with detailed radiative transfer calculations.
This is primarily because detailed radiative transfer leads to frequency 
diffusion, causing some of the original photons closer to the line center
to diffuse out to the wings.
The shift in frequency not only results in smaller scattering 
optical depth but also is accompanied by spatial diffusion, both leading to
larger transmitted flux near the center.

\subsection{Shift in the Peak of \lya Spectra}
\label{sec:peakshift}

\begin{figure*}
\plotone{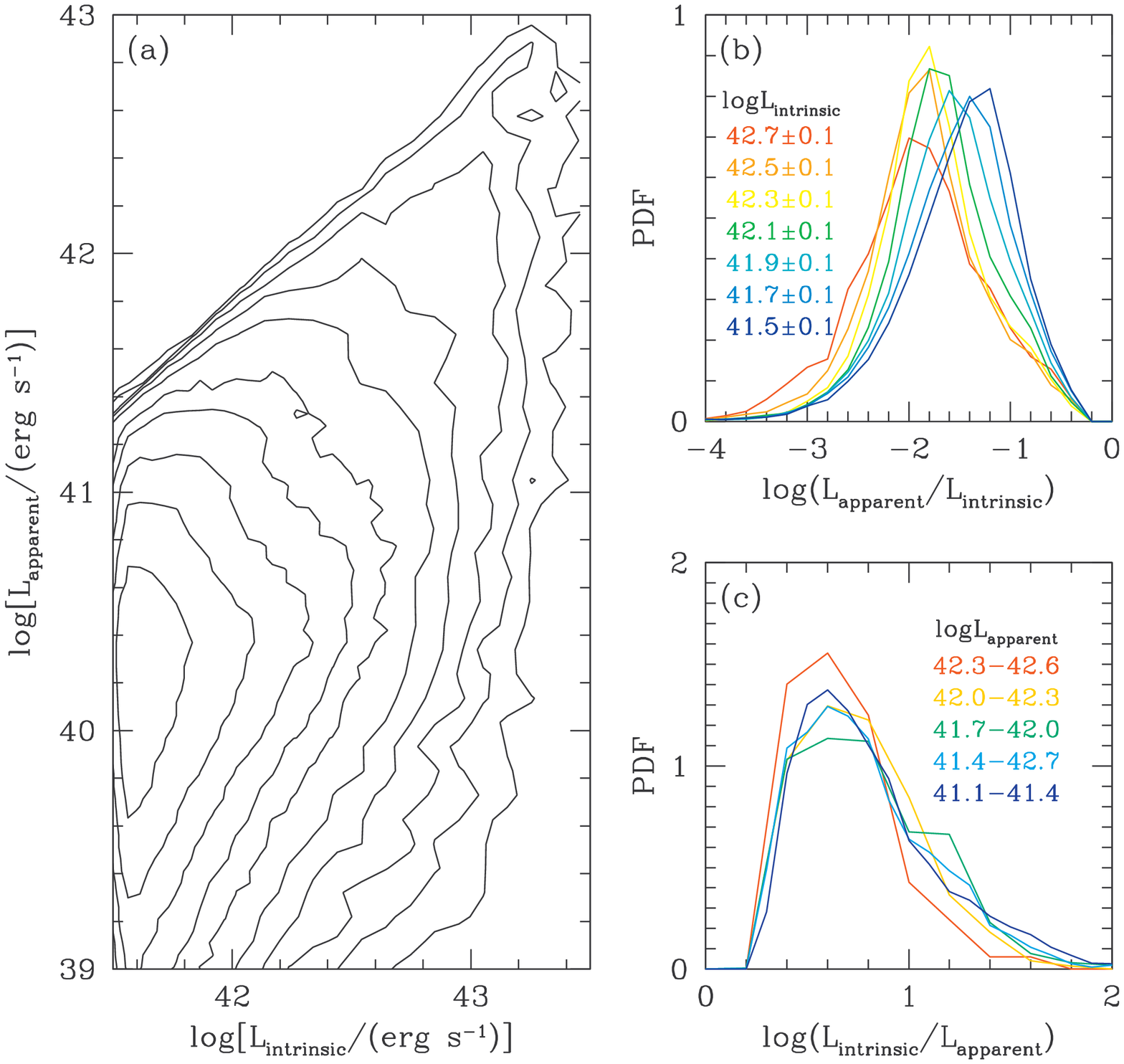}
\caption[]{
\label{fig:LL}
Relation between the observed (apparent) and the intrinsic \lya luminosities
of LAEs. {\it Panel (a)}: The joint distribution of apparent and intrinsic 
\lya luminosities. Adjacent contours differ by a factor of 2 in contour levels.
A vertical cut in this plot gives the probability distribution function (PDF) 
of apparent luminosity at a given intrinsic luminosity [{\it Panel (b)}], 
and a horizontal cut
gives the distribution of intrinsic luminosity at a given apparent luminosity
[{\it Panel (c)}]. Only halos above $5\times 10^9\hMsun$ are considered
in our model, which corresponds to intrinsic luminosity above 
$\sim 10^{41.5} \ergsec$. As a results, the apparent luminosity is complete
above $\sim 10^{41.2} \ergsec$.
}
\end{figure*}

\begin{figure}
\plotone{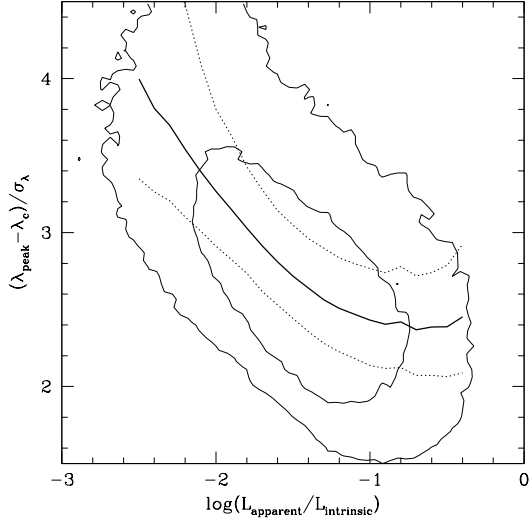}
\caption[]{
\label{fig:Lratio_peakshift}
Correlation between apparent to intrinsic luminosity ratio and wavelength 
shift at peak \lya flux. The peak wavelength shift is in units of the 
intrinsic line width $\sigma_\lambda$, which is set by the virial 
temperature of the host halo in our model. Contours show the correlation
and they enclose the 68\% and 95\% distribution, respectively.
The solid curve is the median wavelength shift as a function of luminosity
ratio. The two dotted curves delineate the upper and the lower quartiles.
}
\end{figure}

\begin{figure*}
\plotone{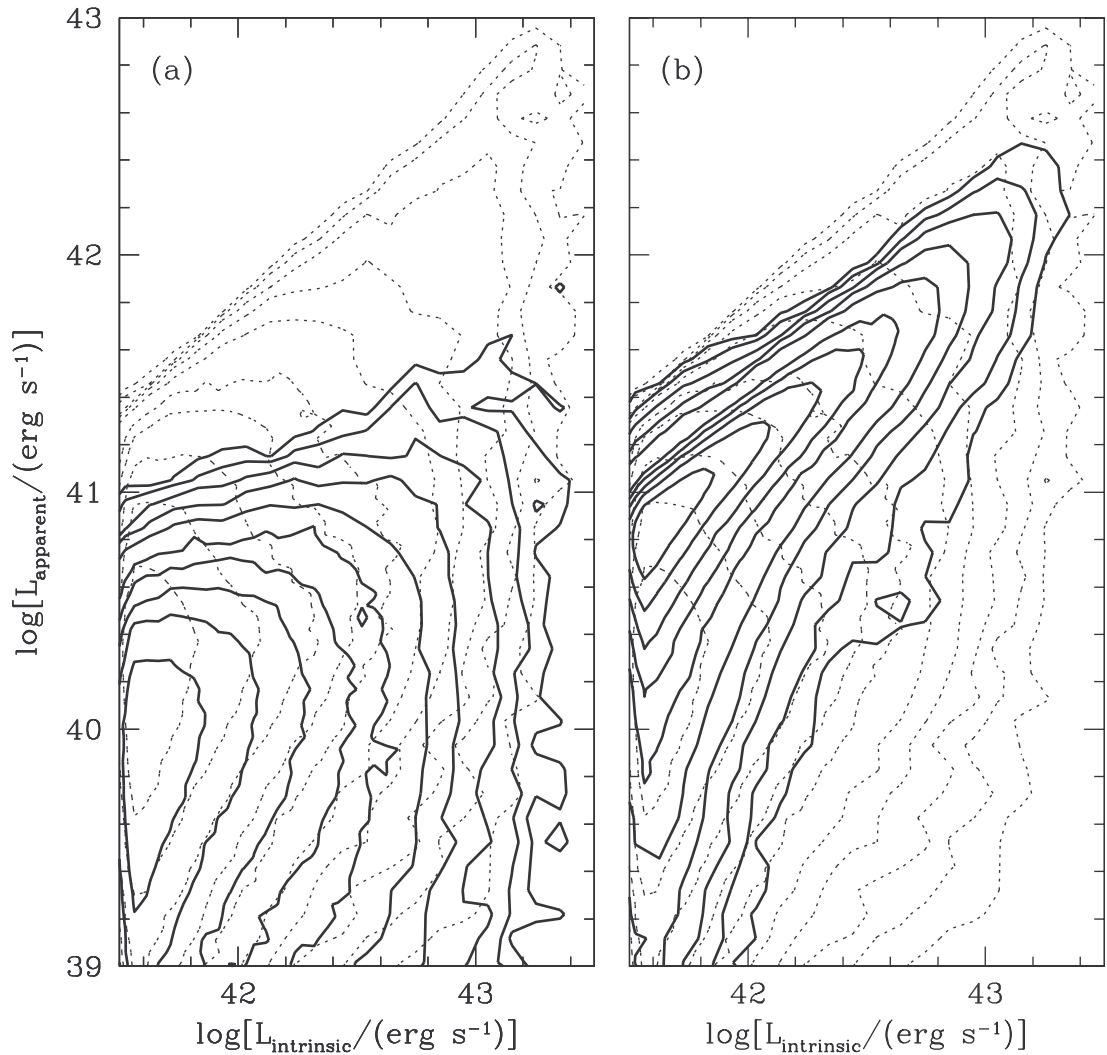}
\caption[]{ 
\label{fig:LL_cmp}
Comparison of the relations between the observed (apparent) and the intrinsic
\lya luminosities from different models. The dotted contours are from our 
model with full calculation of \lya radiative transfer (the same contours as
in Fig.~\ref{fig:LL}$a$). The solid contours in panel ($a$) are from the
simple $\exp(-\tau_\nu)$ model with the initial \lya line width given by
halo virial temperature, as adopted in our model with full radiative transfer
calculation. The solid contours in panel ($b$) are also from the
$\exp(-\tau_\nu)$ model, but the initial \lya line width is determined by
the circular velocity at halo virial radius, which is about 2.3 times larger
than the one adopted in panel ($a$). 
Note that the $\exp(-\tau_\nu)$
model and the full calculation in panel ($b$) assume different initial \lya 
line widths, so it is not an apple-to-apple comparison. It shows that
modifying the initial line width of the $\exp(-\tau_\nu)$ model does not
lead to a result mimicking that from the full calculation.
}
\end{figure*}

\begin{figure}
\plotone{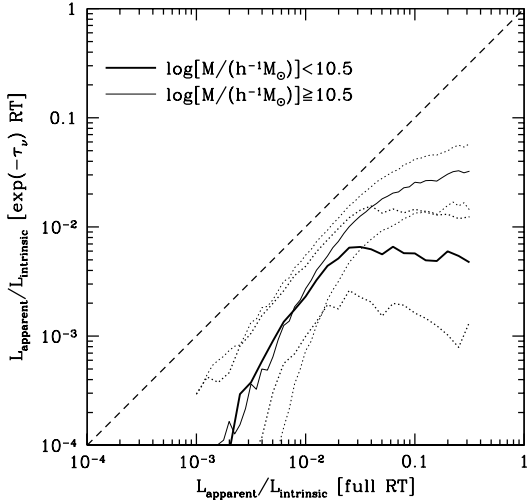}
\caption[]{
\label{fig:one_to_one_cmp}
Comparison between the apparent-to-intrinsic \lya luminosity ratios 
from our full radiative transfer calculation (full RT) and the 
$\exp(-\tau_\nu)$ model.
The thin solid curve shows the median ratio from the $\exp(-\tau_\nu)$ 
model as a function of the ratio from the full RT for halos below 
$10^{10.5}\hMsun$. The two thin dotted curves indicate the lower and upper 
quartiles. The set of thick curves are for halos above $10^{10.5}\hMsun$.
The diagonal dashed line is the line of equality.
}
\end{figure}

\begin{figure}
\plotone{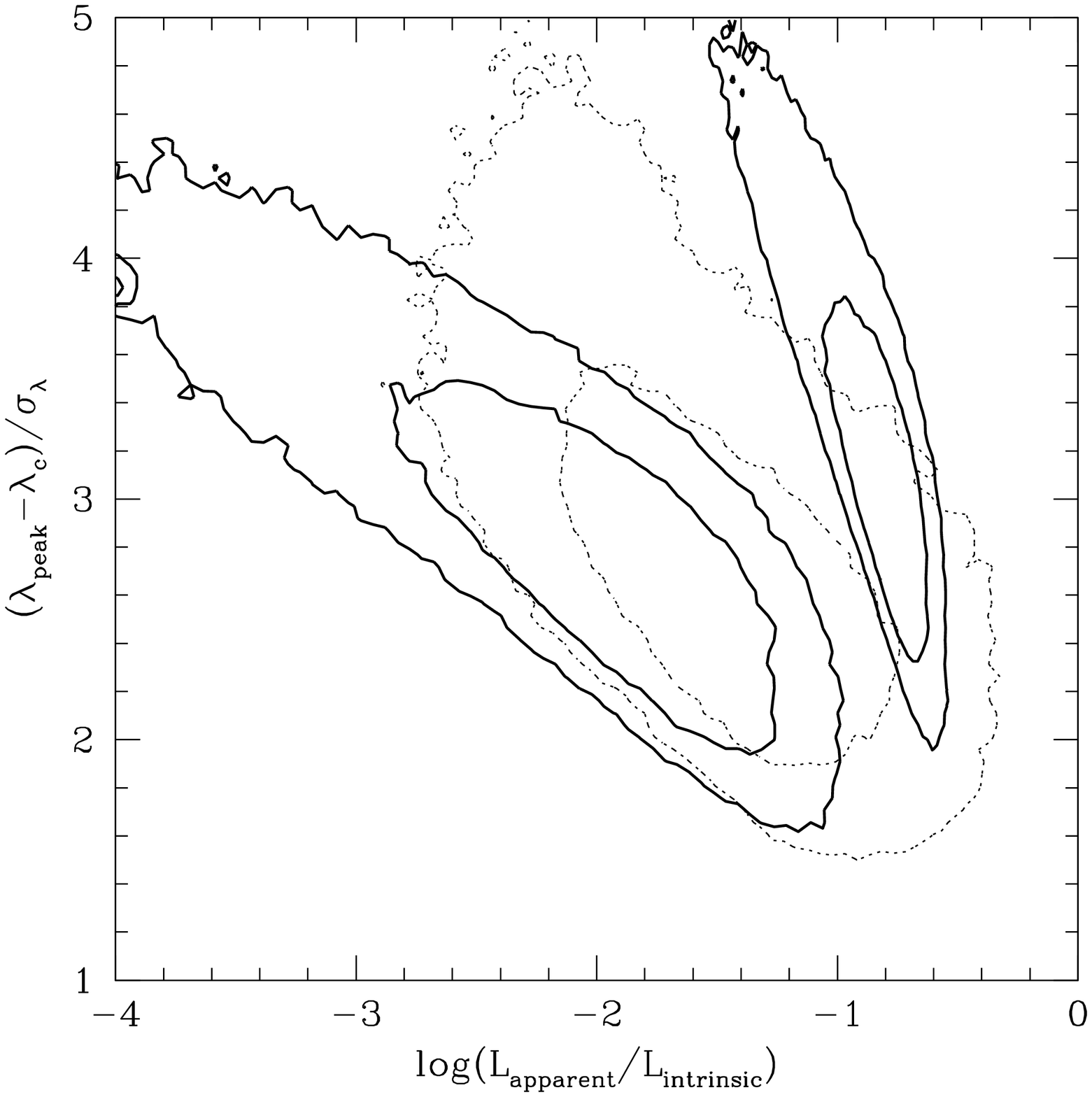}
\caption[]{
\label{fig:Lratio_peakshift_cmp}
Comparison of the correlation between apparent to intrinsic luminosity ratio 
and peak wavelength shift from different models. The dotted contours are from 
our model with full calculation of \lya radiative transfer (the same contours 
as in Fig.~\ref{fig:Lratio_peakshift_cmp}). The solid contours on the left are 
from the simple $\exp(-\tau_\nu)$ model with the initial \lya line width given 
by halo virial temperature, as adopted in our model with full radiative 
transfer calculation. The solid contours on the right are also from the
$\exp(-\tau_\nu)$ model, but the initial \lya line width is determined by
the circular velocity at halo virial radius, which is about 2.3 times larger
than the one from halo virial temperature.
For all the three sets of contours, the peak wavelength shift is in units of 
the line width $\sigma_\lambda$ set by halo virial temperature. 
The two contours in each case show the 68\% and 95\% distribution, 
respectively.
}
\end{figure}

Together with the assumed intrinsic properties, LAEs identified in 
the post-scattering IFU-like data cube from our model enable a statistical 
study of the relations between the observed and intrinsic quantities, which
include spectral features and luminosity.

As shown in Figure~\ref{fig:imgspec_sel}, owing to radiative transfer 
effect, the peak in the observed \lya spectra is at a wavelength longer 
than that in the intrinsic spectra. 
In the left panel of Figure~\ref{fig:peakshift}, we plot the
distribution of the shift as a function of host halo mass. 
The broad distribution of the shift reflects the distribution of 
circumgalactic and intergalactic environments (density and velocity 
structures), which affect the radiative transfer of \lya photons 
(see \S~\ref{sec:factors}).
This apparent wavelength shift in the peak depends on host halo mass,
or intrinsic luminosity of the source given that star formation rate is 
tightly correlated with halo mass in the reionization simulation. 
In general, the distribution is skewed to large shifts. 
For sources in lower mass halos, the distribution is narrower and the average 
shift is smaller. For sources
in halos of $\sim 10^{10}\hMsun$, the median shift corresponds to a velocity 
of $\sim 70 \kms$, while for $\sim 10^{11}\hMsun$ halos, the value is 
$\sim 200 \kms$. 
If the observed \lya emission was from photons backscattered from the far side 
of galactic wind \citep[e.g.,][]{Franx97,Adelberger03}, the above shift would
lead to an overestimate of the wind velocity as long as it is estimated by
the apparent velocity difference between the \lya emission and optical 
emission/absorption lines.
The apparent wavelength shift also translates to a shift in the apparent 
position/redshift of the source along the line of sight (top axis of 
the left panel of Fig.~\ref{fig:peakshift}).
This position shift and its
scatter would result in slight distortion and smoothing in the clustering
of LAEs in redshift space (see Paper II for more details).

In the right panel of Figure~\ref{fig:peakshift}, the apparent shift of
the \lya line peak is put in units of the intrinsic \lya line width 
$\sigma_\lambda$, which is
assumed to be determined by the halo virial temperature. Although the trends
seen in the left panel are still evident, the variation of the distribution 
as a function of halo mass becomes weaker. Roughly speaking, the median shift 
is about $3\sigma_\lambda$ and the scatter is about 0.6--0.9$\sigma_\lambda$.
In terms of velocity, the median shift is approximately 
$100[M_h/(10^{10}\hMsun)]^{1/3}\,\kms$.

The distribution of \lya peak shift we discuss so far is as a function of
intrinsic \lya luminosity (i.e., halo mass). From the point of view of 
observation, it is of great interest to show the distribution as a 
function of the observed \lya luminosity. Figure~\ref{fig:peakshift_obs} 
plots the distribution of \lya peak shift as a function of observed (apparent) 
\lya luminosity. As shown later in \S~\ref{sec:LL}, at fixed intrinsic \lya 
luminosity, the observed (apparent) \lya luminosity has a broad distribution,
and vice versa. The peak shift distribution at fixed observed luminosity is 
therefore contributed by sources residing in halos of a broad range of mass.
We note that the model luminosity in the plot should be increased by about 
0.7 dex to match the $z\sim 5.7$ observation (see \S~\ref{sec:observation}).

\subsection{Apparent and Intrinsic \lya Luminosities}
\label{sec:LL}

The observed \lya flux $F_{{\rm Ly}\alpha}$ from an LAE source in the 
narrow-band image comes from the central part, where the surface brightness 
is high. That is, only a fraction of the extended radiation composed of 
scattered \lya photons can be observed. The observationally inferred \lya 
luminosity $L_{\rm apparent}=4\pi D_L^2 F_{{\rm Ly}\alpha}$ is therefore 
expected to be lower than the intrinsic \lya luminosity $L_{\rm intrinsic}$, 
where $D_L$ is the luminosity distance and isotropic emission is assumed in 
computing the apparent luminosity from the observed flux.

We compare the intrinsic and apparent \lya luminosities in our model. 
Figure~\ref{fig:LL}$a$ gives the joint distribution of $L_{\rm intrinsic}$
and $L_{\rm apparent}$.  From the joint distribution, we can obtain the 
distribution of $L_{\rm apparent}$ at a fixed $L_{\rm intrinsic}$ or vice 
versa, through a vertical or horizontal cut. Since we only model LAEs in halos 
above $5\times 10^9\hMsun$, we are limited to sources with $L_{\rm intrinsic}$
above $\sim 10^{41.5}\ergsec$. When considering the observed luminosity,
sources are complete for $L_{\rm apparent}\ga 10^{41.2}\ergsec$. We note that 
the \lya luminosity limits may change if the assumed IMF and SFR differ from 
those in our model.

In the luminosity range probed by our model, the apparent \lya luminosity
peaks at a few percent of the intrinsic one and is broadly distributed 
(Fig.\ref{fig:LL}$b$). The scatter reflects differences in neutral gas 
distributions (density, velocity, and temperature) around sources of the
same intrinsic luminosity (\S~\ref{sec:factors}). The distribution shifts 
slightly towards higher values for sources of lower intrinsic luminosity. 
If we define the ratio of the apparent to intrinsic luminosity as flux 
suppression, the suppression on average appears to be smaller for 
intrinsically fainter sources, which seems counter-intuitive.
Such a shift is a consequence that the environment of low mass halos
are on average less dense than that of massive halos, and that the environment
is important in shaping the observability, to be discussed in detail in 
\S~\ref{sec:factors}.

From the point of view of observation, it is interesting to ask what the 
observed \lya luminosity implies about the intrinsic one. 
Figure~\ref{fig:LL}$c$ shows the intrinsic luminosity distribution at a given
apparent luminosity. The distributions for different values of 
$L_{\rm apparent}$ are similar in terms of the intrinsic to apparent 
luminosity ratio. In general, the intrinsic luminosity is about 3--12
times the observed luminosity. In other words, a large fraction of the 
escaped \lya photons are invisible. For estimating \lya escape fraction from
observations, this is a systematic factor that 
needs to be taken into account.

We find that the flux suppression is correlated with the shift 
in the peak of \lya profile (\S~\ref{sec:peakshift}),
as shown in  Fig.\ref{fig:Lratio_peakshift}. 
For sources with a larger suppression 
in \lya flux, the peak of the spectra shifts more towards red.
This correlation has only a weak dependence 
on halo mass or intrinsic luminosity and in Figure~\ref{fig:Lratio_peakshift} 
all sources in our model are included. 
Clearly, the correlation is a 
consequence of the radiative transfer: \lya photons diffuse more in frequency 
as they experience more scatterings. 
The correlation is driven by the dependence of the \lya radiative transfer 
on environments, i.e., the circumgalactic and intergalactic density and 
velocity structures (see \S~\ref{sec:factors}). 

It is worth pointing out that the simple $\exp(-\tau_\nu)$ model can only
give qualitatively trends seen in our results. In Figure~\ref{fig:LL_cmp},
we compare the $L_{\rm apparent}$--$L_{\rm intrinsic}$ distribution from
our model of full radiative transfer with those from the $\exp(-\tau_\nu)$ 
model. The $\exp(-\tau_\nu)$ model in Figure~\ref{fig:LL_cmp}$a$ assumes the
same intrinsic \lya line width as in our model, which is set by halo virial
temperature. It is evident that at the same intrinsic \lya luminosity, the 
$\exp(-\tau_\nu)$ model leads to much lower apparent \lya luminosities than
the full calculation. In particular, the suppression from the $\exp(-\tau_\nu)$
model becomes much stronger for sources of higher intrinsic \lya luminosity 
(or halo mass) because of the high density and peculiar velocity. 
The trend is similar to what \citet{Iliev08} find. As they point out, 
peculiar velocity plays an important role in shaping the luminous end
of the observed \lya luminosity function. 
The frequency and spatial diffusions in the full calculation can compensate 
the density and peculiar velocity effect, weakening the 
suppression. 
Compared to the $\exp(-\tau_\nu)$ model, the suppression from the full 
calculation does not become much stronger for sources of higher intrinsic 
\lya luminosity (also see Figure~\ref{fig:one_to_one_cmp}).

The $\exp(-\tau_\nu)$ model in Figure~\ref{fig:LL_cmp}$b$ 
adopts an intrinsic \lya line width 2.3 times that used in 
Figure~\ref{fig:LL_cmp}$a$, which corresponds to the circular velocity
at halo virial radius. This value of intrinsic line width is used in 
some previous work \citep[e.g.,][]{McQuinn07,Iliev08} with the 
$\exp(-\tau_\nu)$ model. This model boosts the apparent \lya luminosity, 
but the suppression is still stronger than the full calculation at high halo 
mass end. In addition, the distribution of apparent \lya luminosity at 
fixed intrinsic \lya luminosity is much narrower than that from the full
calculation. We do not have results from a full radiative transfer calculation 
with the larger intrinsic \lya line width yet, but we expect that the 
difference between such a full calculation and the $\exp(-\tau_\nu)$ model 
is similar to that seen in Figure~\ref{fig:LL_cmp}$a$. 
We caution that 
Figure~\ref{fig:LL_cmp}$b$ does not show an apple-to-apple comparison, since
the $\exp(-\tau_\nu)$ model and the full calculation assume different
intrinsic line widths. Nevertheless, it indicates that
modifying the $\exp(-\tau_\nu)$ by varying the intrinsic line width does 
not lead to a match to the full radiative transfer calculation.

To further see the difference between the full radiative transfer model and
the $\exp(-\tau_\nu)$ model with the same intrinsic \lya line width setup,
we compare the apparent-to-intrinsic \lya luminosity ratios predicted from 
the two models on a one-to-one basis (Figure~\ref{fig:one_to_one_cmp}). 
We summarize the scatter plot by showing the median ratio (solid curve) 
from the $\exp(-\tau_\nu)$ model as a function of the ratio from our model
(the ``true'' ratio), together with the lower and upper quartiles (dotted 
curves). Thin and thick curves are for sources in halos below and above 
$10^{10.5}\hMsun$, respectively. In general, there is a trend that the ratio 
from the $\exp(-\tau_\nu)$ model increases with the ``true'' value and this 
trend seems to break down in massive halos at high values of the ``true'' 
ratio. In a limited range (around $\Lapp/\Lint \sim 10^{-2}$), the median 
ratio from the $\exp(-\tau_\nu)$ appears to be a constant shift from the 
``true'' ratio. However, even if we apply a correction to account for the 
shift, the $\exp(-\tau_\nu)$ model would underpredict the luminosity ratio 
outside of the above narrow range, in particular towards higher values of 
the ``true'' ratio. Even within the narrow range, the ratio from the 
$\exp(-\tau_\nu)$ model has a large scatter (a factor of a few) at a 
fixed ``true'' ratio.

The \lya flux suppression in our $\exp(-\tau_\nu)$ model appears to be much
stronger than seen in previous work \citep[e.g.,][]{McQuinn07,Iliev08}. 
Compared to our work, previous work assumes a much wider initial \lya line
width and lower gas temperature (set to be $10^4$K). The differences in the
above two factors largely explains the differences in the results. For more
detailed explanations and discussions, see Appendix~\ref{sec:expmtau_test},
where we perform several tests with the $\exp(-\tau_\nu)$ model by varying
the initial \lya line width and gas temperature.

Figure~\ref{fig:Lratio_peakshift_cmp} compares the correlation between 
apparent to intrinsic luminosity ratio and peak wavelength shift from
our full radiative transfer model (dotted contours) and the above two 
$\exp(-\tau_\nu)$ models (solid contours). Although the sign of the correlation
is the same for all three models, the $\exp(-\tau_\nu)$ model
with smaller (larger) intrinsic \lya line width gives a smaller (larger) 
slope in the correlation than the full calculation. 

The comparisons with the $\exp(-\tau_\nu)$ model results in this section 
demonstrate that the $\exp(-\tau_\nu)$ model can provide a qualitative 
understanding of the results, but there is no simple way to modify the results 
from the $\exp(-\tau_\nu)$ model to match those of the full calculation. Full 
radiative transfer calculation is necessary to 
obtain quantitatively correct results in \lya emission properties of LAEs.

\subsection{\lya Luminosity Function of LAEs}
\label{sec:LF}

An important product of surveys of LAEs is the \lya LF, 
one of the most widely studied statistical properties of LAEs. The LF can 
be used to infer the relation between LAEs and their host dark matter halos. 
The evolution of LFs around the reionization epoch can 
probe the reionization of the universe \citep[e.g.,][]{Malhotra04,Haiman05}.
As we have shown in \S~\ref{sec:LL},
the observed \lya luminosity of LAEs differ from the intrinsic one.
Here we study the \lya LF of LAEs from our model. 

\begin{figure}
\plotone{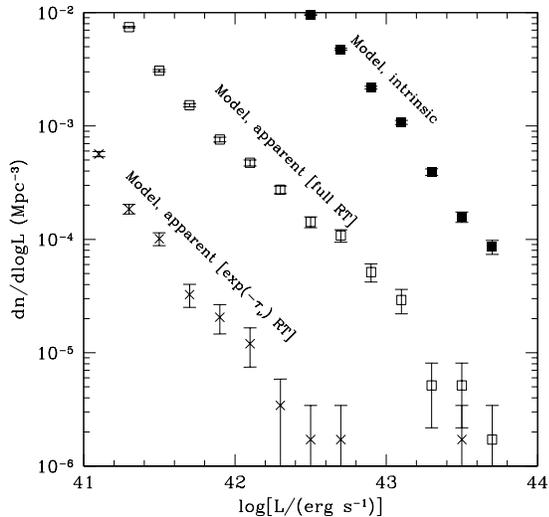}
\caption[]{
\label{fig:LF}
\lya luminosity function of $z\sim 5.7$ LAEs.
The intrinsic luminosity function in our model is represented by filled 
squares. The apparent luminosity function from a full radiative transfer 
calculation is plotted as open squares. For comparison, the crosses show
the apparent luminosity function with a simple $\exp(-\tau_\nu)$ radiative
transfer treatment. Poisson errors are plotted. 
}
\end{figure}

The star formation prescription adopted in the reionization simulation 
leads to a tight correlation between star formation rate and halo mass.
Therefore the intrinsic \lya luminosity is tied to the halo mass and the 
intrinsic \lya LF largely reflects the halo mass function. 
In Figure~\ref{fig:LF}, the filled squares show the intrinsic \lya LF,
$\Phi_i(L_{\rm intrinsic})$. 
The apparent \lya LF, $\Phi_a(L_{\rm apparent})$, is related to the intrinsic one through
\begin{equation}
\Phi_a(L_{\rm apparent})=\int_0^\infty p(L_{\rm apparent}|L_{\rm intrinsic})\Phi_i(L_{\rm intrinsic}) dL_{\rm intrinsic},
\end{equation}
where $p(L_{\rm apparent}|L_{\rm intrinsic})$ is the probability density
of the apparent luminosity at a given $L_{\rm intrinsic}$ 
(Fig.~\ref{fig:LL}$b$). 
The apparent \lya LF can be directly read off from
Figure~\ref{fig:LL}$a$ and is shown as open squares in Figure~\ref{fig:LF}.
In terms of luminosity, the apparent \lya LF shifts towards the faint end by a
factor of 5--20 with respect to the intrinsic one and the shift is larger 
at the faint end. Formally, the intrinsic LF can be inferred from 
the observed one by
\begin{equation}
\Phi_i(L_{\rm intrinsic})=\int_0^\infty p(L_{\rm intrinsic}|L_{\rm apparent})\Phi_a(L_{\rm apparent}) dL_{\rm apparent},
\end{equation}
where $p(L_{\rm intrinsic}|L_{\rm apparent})$ is the probability density
of the intrinsic luminosity at a given $L_{\rm apparent}$ 
(Fig.~\ref{fig:LL}$c$). 
However, even if 
$p(L_{\rm intrinsic}|L_{\rm apparent})$ is known or assumed {\it a priori},
it is not enough to infer the intrinsic LF from the observed one. The reason
is that at a given intrinsic luminosity, the distribution of apparent 
luminosity can have a tail to low value (Fig.~\ref{fig:LL}). So one needs to 
have observations of the apparently faint LAEs to recover
the full information of the intrinsic luminosity distribution, or one has to
rely on the extrapolation of the observed LF to the faint end.
 
For comparison, the crosses in Figure~\ref{fig:LF} show the \lya LF from the
$\exp(-\tau_\nu)$ model, which adopts the same intrinsic \lya line width as
the full radiative transfer calculation. As already demonstrated in 
\S~\ref{sec:LL}, the \lya flux is more strongly suppressed in the 
$\exp(-\tau_\nu)$ model. The resultant apparent \lya LF looks like the 
intrinsic one shifting toward the faint end by more than two orders of 
magnitude in luminosity. With the same intrinsic \lya LF, the apparent
\lya LF from the full radiative transfer model is significantly higher
than that from the $\exp(-\tau_\nu)$ model, a consequence
of the frequency and spatial diffusion of \lya photons from scatterings.

\section{Implications for Observed Properties of LAEs}
\label{sec:observation}

Our model of LAEs predicts the relation between observed \lya emission and the
intrinsic one. With simple assumptions about the intrinsic properties of LAEs,
we are able to predict an array of observational properties of LAEs. In this
section, we compare our model predictions to observations for $z\sim 5.7$ LAEs
and attempt to understand various observed properties of LAEs. We focus 
here on the \lya LF, the UV LF, and the 
\lya equivalent width (EW) distribution of LAEs.

\subsection{\lya Luminosity Function}
\label{sec:LFlya}

\begin{figure}
\plotone{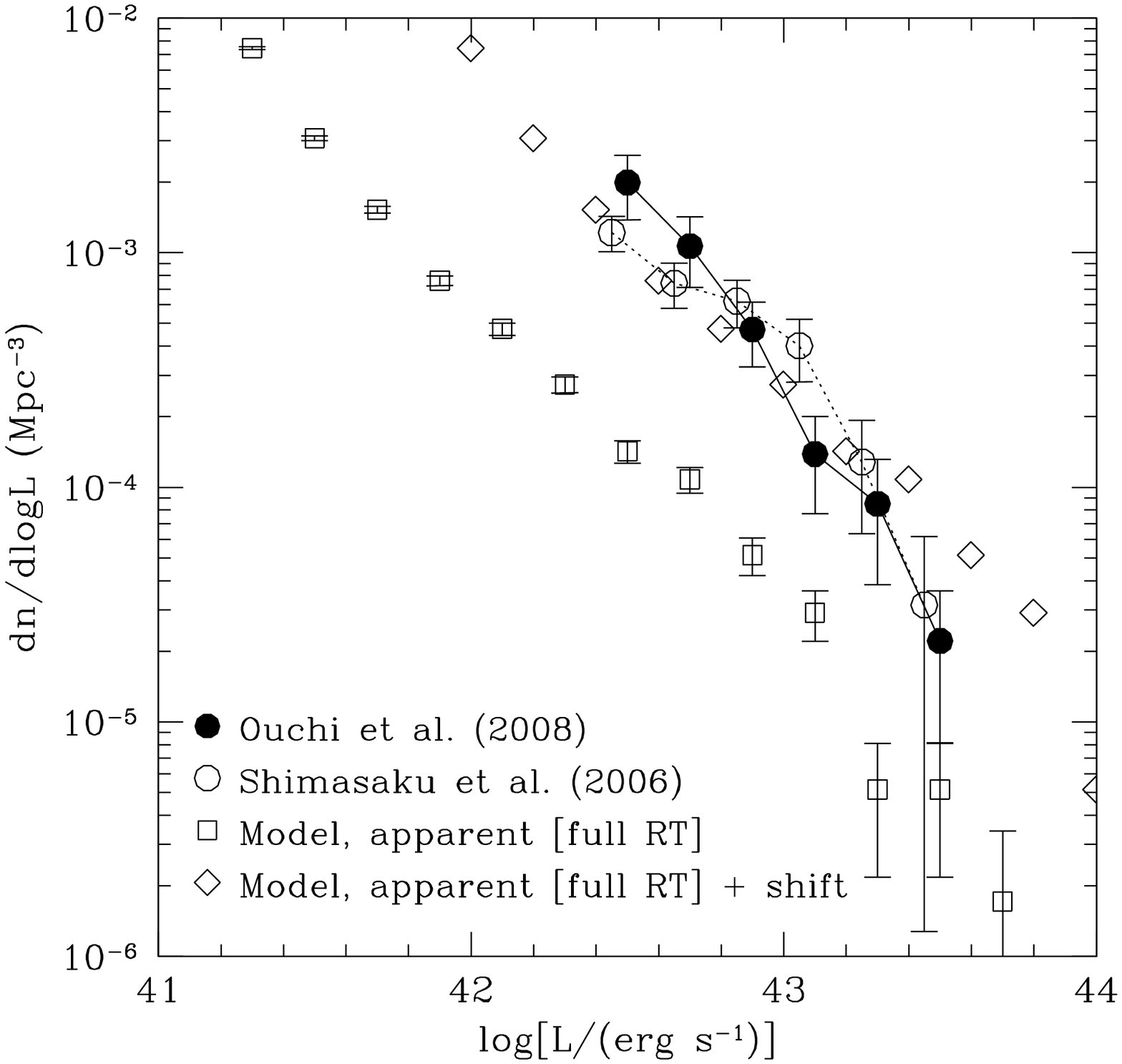}
\caption[]{
\label{fig:LFcmpobs}
Observed \lya LF of $z\sim 5.7$ LAEs. 
Solid circles connected by solid lines are the observed LF 
for $z\sim 5.7$ LAEs in the 1 deg$^2$ SXDS \citep{Ouchi08}. 
Open circles connected by dotted lines are the observed one from the 0.2 
deg$^2$ Subaru Deep Field (a different sky area from SXDS; 
\citealt{Shimasaku06}).
Open squares are the apparent LF from our model with a full 
radiative transfer calculation, and diamonds are the same but with the \lya 
luminosity boosted by a factor of 5 (see text for details).
}
\end{figure}
 
In our model, the apparent \lya luminosity corresponds to the observed one.
The open squares in Figure~\ref{fig:LFcmpobs} (the same as in Fig.~\ref{fig:LF})
show the predicted \lya LF of LAEs from the model. Filled circles are the 
measurement based on 401 LAEs from the 1 deg$^2$ SXDS \citep{Ouchi08}, while 
open circles show that based on 89 LAEs in a 0.2 deg$^2$ Subaru Deep Field 
\citep{Shimasaku06} that covers a different sky area from SXDS. 

Compared with the observed \lya LF, the apparent LF from our model appears 
to be one order of magnitude lower in {\it normalization}. This seems to put 
into question our model. However, the {\it normalization} is not necessarily 
a good indicator of the importance of the difference. A more sensible way is 
to characterize the difference by the shift in luminosity scale --- with 
respect to the observed \lya LF, the apparent LF from our model is shifted 
by a factor of 3--6 to the left (toward the low luminosity end). 
If we simply increase the apparent \lya luminosity by a factor of 5, the
apparent LF (shown as diamonds in Fig.~\ref{fig:LFcmpobs}) shows a better 
match to the observed one. The factor of 3--6 discrepancy between the model
and the data implies that we are missing some physics in our modeling. 
For comparison, the apparent \lya LF based on the simple $\exp(-\tau_\nu)$ 
model (crosses in Fig.~\ref{fig:LF}) corresponds to about two orders of 
magnitude shift in luminosity with respect to the observed one. Again, the 
frequency and spatial diffusion from the realistic radiative transfer in 
our model enhances the probability of being detected for \lya photons. 

Before discussing possible solutions to the factor of 3--6 shift in 
luminosity, we justify that changing the {\it normalization} of \lya LF 
cannot be the solution for the discrepancy. The normalization can only be 
changed by changing the amplitude of the halo mass function. However, we do 
not expect a large uncertainty in the latter with the current constraints on 
cosmology. The reionization simulation that our model is based on adopted 
cosmological parameters in accordance with WMAP five-year results. To estimate 
the cosmology-caused change in the halo mass function, we make use of the 
analytic formula given by \citet{Sheth99}. Not surprisingly, the uncertainty 
in $\sigma_8$ dominates the amplitude change in the halo mass function. 
Within the 1-$\sigma$
uncertainties of the WMAP five-year cosmological parameters \citep{Dunkley09},
the cosmological parameters used in the simulation already put the halo
mass function amplitude in the high end. Different combinations of 
cosmological parameters within their 1-$\sigma$ uncertainty ranges can only 
boost the amplitude by $\sim 30\%$. Therefore, we conclude that 
cosmology-caused amplitude change in halo mass function does not help much 
in making our model \lya LF match the observed one.

\lya luminosity depends on the amount of massive stars that emit ionizing 
photons. The conversion from star formation rate to intrinsic \lya luminosity 
adopted in our model assumes \citet{Salpeter55} IMF. For a fixed total 
stellar mass with
star mass distributed in the range of 0.1--100$\Msun$, simply changing to a 
\citet{Chabrier03} IMF, which has shallower slopes than the Salpeter IMF below
a characteristic mass of 1$\Msun$, can increase the ionizing photons by a 
factor of $\sim$1.6. This assumes no evolution in the stellar IMF with cosmic 
time. Recent studies (e.g., \citealt{Dave08,vanDokkum08}) show evidence of an
evolving IMF and observations of $z\sim$7 dropout galaxies also implies
an IMF changes towards high redshift \citep[e.g.,][]{Ouchi09}. 
The form of evolution can be thought as an increasing 
characteristic mass in the Chabrier IMF with increasing redshift. By adopting
either form of the IMF at high redshift proposed by \citet{Dave08} and 
\citet{vanDokkum08}, we find an increase in the ionizing photons by a factor
of $\sim$5.5 with respect to the Salpeter IMF. Therefore, for the instantaneous
star formation rate, the \lya luminosity could have the amount of enhancement
we need for the model prediction to match the observation. Obviously, the 
situation is not as simple as this and there are additional factors, such as 
the dependence on star formation history and metallicity. 
Overall, a possible factor of 3--6 in underestimating the intrinsic \lya luminosities 
in our present model may be physically accounted for 
and that would put our model in accord with observations.

\begin{figure}
\plotone{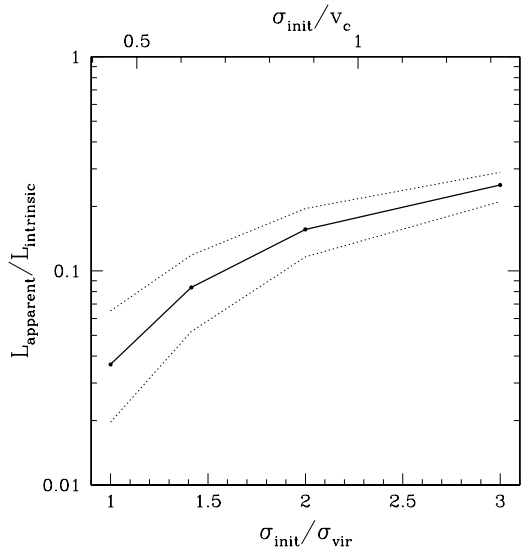}
\caption[]{
\label{fig:Lratio_Width}
Effect of the initial \lya line width. Plotted here is the ratio of the 
apparent (observed) to intrinsic \lya luminosity of $z\sim 5.7$ LAEs as 
a function of the initial \lya line width $\sigma_{\rm init}$. The bottom
axis marks the line width in terms of the width $\sigma_{\rm vir}$ set 
by halo virial temperature, while the top axis in units of the circular 
velocity $v_c$ at halo virial radius. The solid and dotted curves are the
median and quartiles of the luminosity ratio distribution. This test is 
done with a small simulation (box size of 25$\hMpc$ on a side). The
distribution is calculated from all sources with halo mass above 
$5\times 10^9\hMsun$. See text.
}
\end{figure}

\begin{figure*}
\plotone{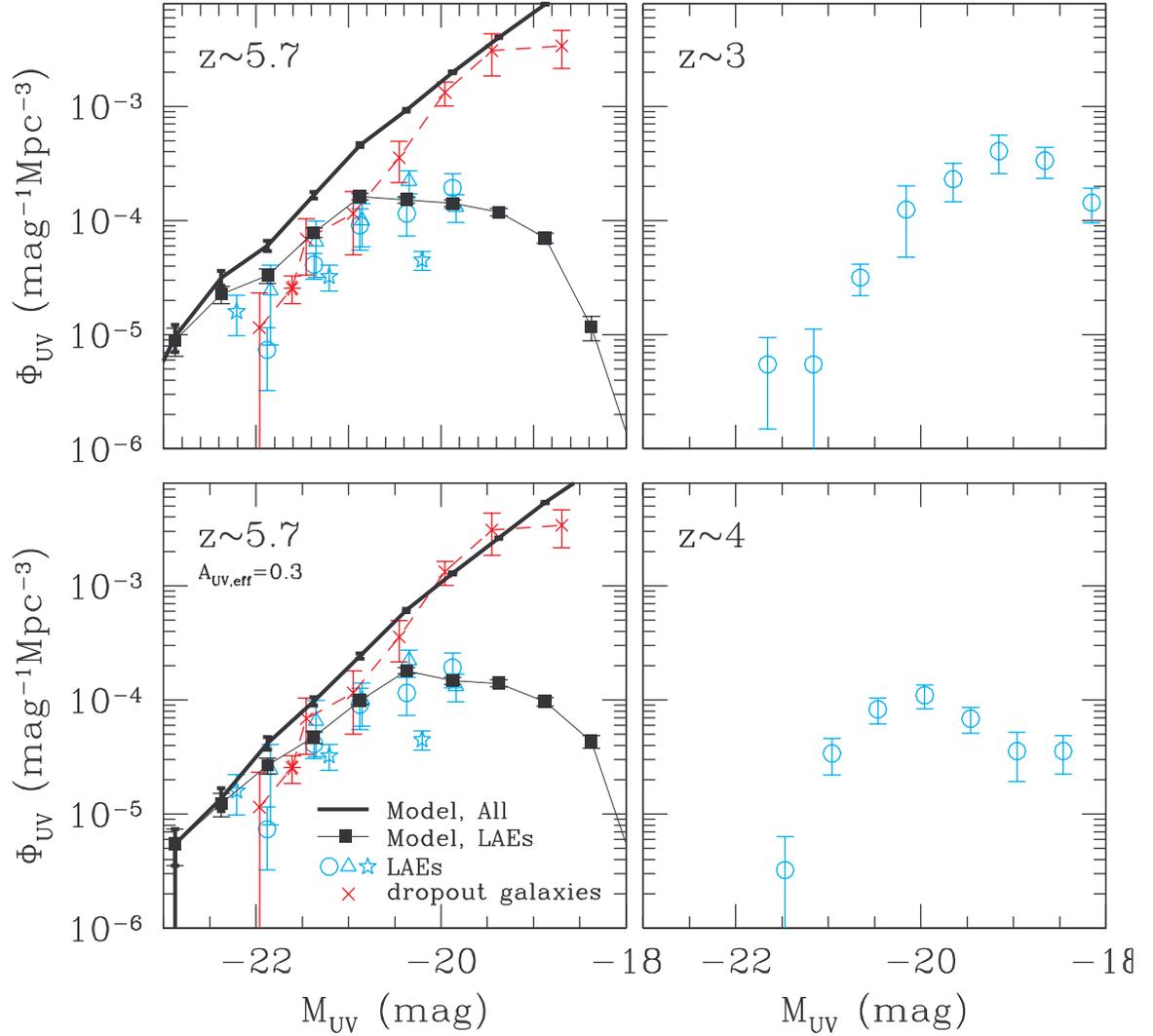}
\caption[]{
\label{fig:uvLF}
UV LFs of LAEs. Left panels are for $z\sim 5.7$ LAEs.
The thick solid curve is the UV LF for all galaxies in our model. The filled 
squares connected by the thin solid curve are the UV LF of model LAEs with the 
apparent \lya luminosity above a threshold. The threshold is chosen so that 
the number density matches that of SXDS $z=5.7$ LAEs (see text). The observed 
UV LFs of LAEs are plotted as open symbols, obtained by \citet{Ouchi08} 
(circles), \citet{Shimasaku06} (triangles), and \citet{Hu06} (stars), 
respectively. The LF of $i$--dropout galaxies at $z\sim 6$ is shown as
crosses \citep{Bouwens06} and asterisk \citep{Shimasaku06}. 
In the bottom left panel, an effective UV extinction of 
$A_{\rm UV,eff}=0.3$ is applied to the two curves from the model (see text).
The two right panels are the observed UV LF for LAEs at $z\sim 3$ and 
$z\sim 4$ in SXDS, taken from \citet{Ouchi08}.
}
\end{figure*}

In addition, a possible increase in the intrinsic \lya width may also increase
the apparent luminosity. In our model, the intrinsic \lya line profile is 
assumed to be Gaussian with width determined by the halo virial temperature. 
The virial temperature is defined as $T_{\rm vir}=GM_h\mu m_H/(3kR_{\rm vir})$ 
\citep{Trac07}, resulting in a line width of $(\mu/3)^{1/2}v_c\simeq 0.44v_c$.
Here $\mu\sim 0.59$ is the mean molecular weight and 
$v_c\simeq 155[M_h/(10^{11}\hMsun)]^{1/3}\kms$ the circular velocity at the 
virial radius. Such an assumption on the line width is conservative. Many 
processes can alter the intrinsic line profile. Disk rotation can change the 
line profile and broaden it. Galaxy merging and galactic wind from star 
formation can also substantially broaden the line. Many authors have adopted 
line width $\sim v_c$ or larger in computing \lya transmission with the 
$\exp(-\tau_\nu)$ model (e.g., \citealt{Santos04,Dijkstra07a,McQuinn07,
Iliev08}). We have tested the effect of the initial line width by performing 
\lya radiative transfer for LAEs in a small simulation of box size of 
25$\hMpc$ as well as for a subset of LAEs in the 100$\hMpc$ box simulation. 
Figure~\ref{fig:Lratio_Width} shows the apparent to intrinsic \lya luminosity
ratio as a function of the initial \lya line width from the simulation with
the 25$\hMpc$ box, inferred from all sources with halo mass above 
$5\times 10^9\hMsun$. Changing the line width we adopt to $v_c$ can lead to
a factor of $\sim$5 increase in the median luminosity ratio. Therefore, the 
effect of initial \lya line width can potentially shift the apparent 
\lya LF by a factor of a few towards the bright end.

Moreover, the star formation prescription adopted in the simulation 
has a number of assumptions and uncertainties
because of the limitation in our understanding
of the baryon-related processes. A higher SFR than that in the simulation 
would lead to an increase in the intrinsic \lya luminosity, and hence an 
increase in the apparent \lya luminosity.

We see that a combination of changes in stellar IMF, intrinsic line width, 
and SFR can solve the problem of a factor of a few mismatch between the 
apparent \lya LF of our model and the observed one. Changing the IMF and SFR 
would also change the reionization process and the neutral fraction of IGM, 
although they are degenerate with uncertain escape fraction of ionizing 
photons. The prediction of \lya luminosity is coupled with
the evolution of the gas ionization state. A self-consistent calculation
is possible, but it is out of the scope of this paper. An accurate estimation 
of the initial \lya profile and width would require detailed calculation of 
the \lya transfer through realistic ISM of high redshift 
galaxies, which is little constrained and difficult to compute from first 
principles presently. We conclude that, owing to model uncertainties, the 
discrepancy between model LF and observed LF is not as serious a problem as 
it appears to be and may be indications of some interesting physics that is 
not considered in our current calculation or the need to have more accurate 
prescriptions of some processes.

\subsection{UV Luminosity Function}
\label{sec:uvLF}

As mentioned in \S\ref{sec:LL}, at a fixed intrinsic \lya luminosity of LAEs, 
the apparent (observed) \lya luminosity has a broad distribution, and vice
versa. Since the intrinsic \lya luminosity is directly correlated with the UV
luminosity, our results mean that the UV LF of the observed LAEs (with apparent 
\lya luminosity above certain threshold) must differ from the intrinsic one.
In what follows, we show the differences caused by the \lya selection and
compare the model UV LF of LAEs with those from observations.

We convert the SFR in halos to the UV luminosity $L_{\rm UV}$ (at 1500\AA) 
through 
\begin{equation}
L_{\rm UV}=8\times 10^{27} [{\rm SFR}/(\Msun {\rm yr}^{-1})]\, 
\ergsec{\rm \,Hz^{-1}},
\end{equation}
which assumes Salpeter IMF and solar metallicity \citep{Madau98}. Following
observers, we express the UV luminosity in AB magnitude, $M_{\rm UV}
=-2.5\log[L_{\rm UV}/(4\pi d_0^2)]-48.60$ with $d_0=10$pc.

The thick solid curve in the top-left panel of Figure~\ref{fig:uvLF} is the 
UV LF for all sources (galaxies) in our model. Because of the tight correlation 
between SFR 
and halo mass, the curve is basically a transformation of the halo mass 
function with a constant mass-to-light ratio. This UV LF is from all the
galaxies in our model, regardless of the apparent \lya luminosity. To be
detected as LAEs, the apparent \lya luminosity should be high, which imposes
a selection function onto the full UV LF. The UV LF of LAEs is from sources
with apparent \lya luminosity above a threshold, which is set in our model
such that the number density of the selected LAEs
(that can be observed) matches that of the $z=5.7$ LAEs in SXDS (about 
$4.0\times 10^{-4} {\rm Mpc}^{-3}$ for our adopted cosmology). The filled
squares connected by thin solid curve shows the UV LF for the LAEs selected
this way. For a cut in apparent \lya luminosity, sources with a higher 
intrinsic \lya luminosity (hence a higher UV luminosity) have a higher
possibility to be selected (see Fig.\ref{fig:LL}). As a result, the UV LF 
of LAEs is close to the full UV LF at the high luminosity end. However, 
in lower mass halos (sources with lower UV luminosity), fewer sources 
can have apparent \lya luminosity high enough to be detected. The UV LF
of LAEs becomes lower than the full UV LF. The ratio of the UV LFs of LAEs
and all galaxies as a function of UV luminosity is nothing more than a 
reflection
of the distribution of apparent \lya luminosity as a function of halo mass 
(Fig.\ref{fig:LL}). As a consequence of this simple effect, the UV LF of 
LAEs becomes flattened as the UV luminosity decreases and drops rapidly towards
the low luminosity end (Fig.\ref{fig:uvLF}).

The predicted features in the UV LF of LAEs are indeed seen in observations.
The cyan points are observed UV LFs of $z=5.7$ LAEs (\citealt{Shimasaku06,
Hu06,Ouchi08}). The LF becomes flat for $M_{\rm UV}>-20.5$. The $z=5.7$ 
observations are not deep enough to show the predicted drop of the UV LF at 
the faint end, but the drop can be clearly seen in the $z\sim3$ and $z\sim 4$ 
data, as shown in the right panels of Fig.\ref{fig:uvLF} (also see Fig.22 of 
\citealt{Ouchi08}). Without any adjustment, the UV LF of LAEs from our model 
is in quite a reasonable agreement with observations. The observed flattening 
of the LF toward lower luminosity is well explained by our model. The 
agreement improves by adding an effective UV extinction of $A_{\rm UV,eff}=0.3$ 
to our model curve (lower left panel of Fig.\ref{fig:uvLF}).
As implied in \S~\ref{sec:LFlya}, the assumed IMF and the adopted SFR in the 
simulation may not be accurate, and we may need a higher UV luminosity. 
Therefore, the effective UV extinction here should be understood as a 
combination of the model uncertainty and dust extinction. 

\citet{Kabayashi10} present an LAE model with \lya escape fraction and UV 
extinction being functions of metal column density and starforming and outflow 
phases of galaxies, which roughly reproduce the observed UV LF. The 
semi-analytic model of \citep{Samui09} with constant \lya escape fraction 
and UV extinction fails to reproduce the turnover of the UV LF towards low 
luminosity end. By contrast, there is no \lya escape fraction parameter
and mass dependent UV extinction in our model, and the radiative transfer 
is the single factor responsible to convert the intrinsic \lya emission
to the observed one. In other words, the \lya escape fraction, defined
as the ratio of apparent to intrinsic \lya luminosity is an output of the 
model. It is encouraging that our model, by accounting for simple physics,
is able to reproduce the features in the observed UV LF. This is an 
independent output of our model and clearly lends credence to our model.

The sources that are not detectable as LAEs because of a low apparent 
\lya luminosity can be detected as galaxies through the dropout technique 
\citep[e.g.,][]{Bouwens03}. In Figure~\ref{fig:uvLF}, the observed UV LF of 
the $z\sim 6$ $i$-dropout galaxies is shown as red crosses \citep{Bouwens06} 
and asterisk \citep{Shimasaku06}. At the faint end, it agrees with the full 
UV LF from our model. It falls slightly steeper than our model curve at the 
bright end. Note that the $i$--dropout technique can miss LAEs with strong 
\lya emission \citep{Ouchi08}. The face values of UV LFs from our model and 
the data suggest that the sum of UV LFs of LAEs and dropout galaxies makes 
the full UV LF. Obviously, the sum should not double-count those LAEs 
that are also detected as dropout galaxies. Observations show that about 
30--50\% of dropout galaxies are detected as LAEs 
\citep[e.g.,][]{Dow-Hygelund07,Stanway07,Ouchi08} (our model implies that this 
fraction can depend on UV luminosity). Roughly accounting for this fraction, 
the nonduplicated sum of the observed UV LFs of LAEs and $i$--dropout galaxies
seems to be in a reasonable agreement with the full UV LF from the model.
This is yet another independent output of our model that agrees with 
observations.

We see that, as a consequence of \lya radiative transfer, LAEs are sources 
that have a strong \lya selection imposed. This selection effect nicely 
explains the shape of the observed UV LF of LAEs and that of the dropout
galaxies.  

\subsection{Distribution of \lya Equivalent Width}
\label{sec:EW}

\begin{figure}
\plotone{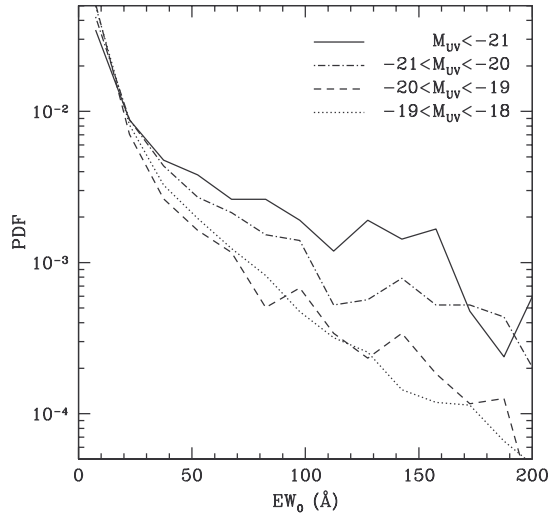}
\caption[]{
\label{fig:EW_PDF}
Distribution of rest-frame \lya equivalent width (EW$_0$) from the model.
Different line types correspond to different UV magnitude.
}
\end{figure}

\begin{figure*}
\plotone{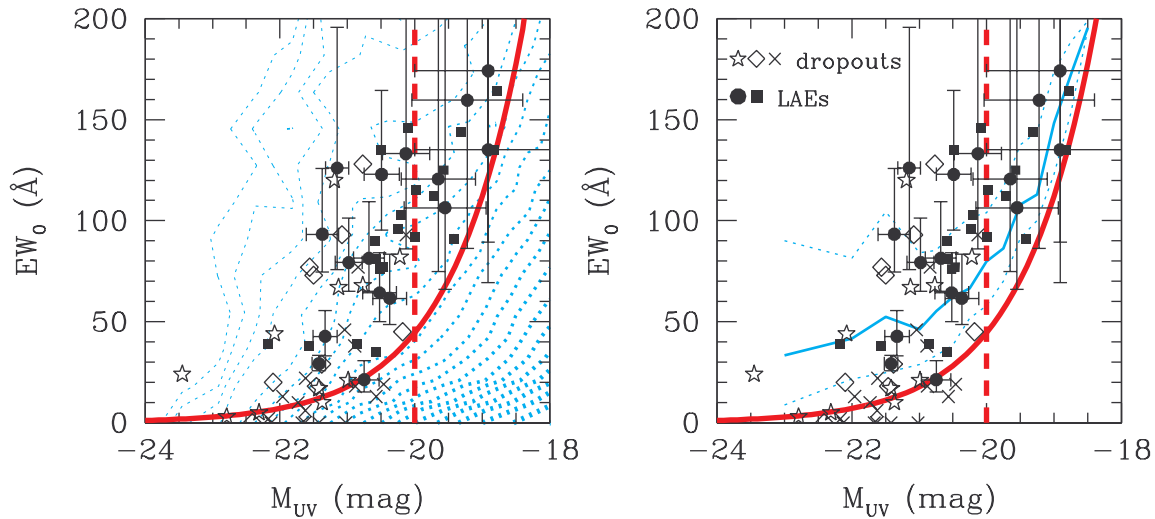}
\caption[]{
\label{fig:EW_conqua}
Rest-frame \lya equivalent width (EW$_0$) as a function of UV luminosity.
In both panels, the thick solid curve corresponds to the threshold of the 
observed (apparent) \lya luminosity for the $z\sim 5.7$ LAEs in 
\citet{Ouchi08}, and the thick dashed vertical line shows the 3--$\sigma$ limit 
for the UV photometry in \citet{Ouchi08}.
Dotted curves in both panels show the distribution of objects in the plane of 
EW$_{0}$ and UV luminosity from our model (see text): in the left panel, the 
contours denote the probability density of objects in the plane (thicker 
contours for higher densities); in the right panel, we plot the median 
(thin solid curve) and quartiles (dotted curves) of EW$_0$ for the
LAEs that can be detected in \citet{Ouchi08}, i.e., for sources above the
thick solid curve.
The data points in both panels are taken from \citet{Ouchi08}: the filled 
circles and squares are for $z\sim 5.7$ LAEs in \citet{Ouchi08} and 
\citet{Shimasaku06} from the Subaru fields; the stars and diamonds represent 
$z\sim 6$ dropout galaxies in \citet{Stanway07} and those compiled by 
\citet{Ando06}; the crosses denote $z\sim 5$ dropout galaxies in 
\citet{Ando06}. 
}
\end{figure*}

As shown in the above two subsections, with simple assumptions to account
for the model uncertainty, the model is able to reproduce the \lya LF and UV 
LF of LAEs. We can go beyond the two LFs to study the joint distribution of
\lya and UV luminosities, which can be casted as the distribution of rest-frame
\lya EW as a function of UV luminosity. This distribution 
is not limited to LAEs and it can include that from the dropout galaxies.

Observationally, it is found that $z\sim 6$ galaxies seem to show a deficit 
of large EW values for UV luminous objects \citep{Ando06,Shimasaku06,Stanway07,
Ouchi08} (see the data points in Fig.~\ref{fig:EW_conqua}). The threshold UV 
luminosity for the deficiency is $M_{\rm UV}=-21.5$ to -21.0 \citep{Ando06}. 
Similar trend is seen for $z\sim$3--5 LAEs as well 
\citep[e.g.,][]{Ouchi08,Shioya09}. The trend is also reported for dropout 
galaxies at different redshifts \citep[e.g.,][]{Shapley03,Ando07,Kajino09,
Pentericci09,Vanzella09}. \citet{Ando06} and other authors invoked the 
differences in dust extinction, amount of internal and surrounding neutral 
hydrogen gas, age of stellar population, and/or gas kinematics between UV 
faint and luminous galaxies as possible causes of the trend seen in \lya 
EW and UV luminosity. \citet{Mao07} present a model of high redshift galaxies
including chemical evolution and dust attenuation. \citet{Kabayashi10} 
also present a semi-analytic model of LAEs, in which \lya and UV are 
attenuated differently by clumpy dust distribution. Both models seem to 
explain the deficiency of high \lya EW in luminous galaxies essentially 
by means of halo mass dependent dust content. On the other hand, based upon 
a pure statistical analysis with $z\sim 3$ galaxies, \citet{Nilsson09} 
conclude that there is no dependence of \lya EW on UV luminosity for LAEs 
and Lyman break galaxies (LBGs). They interpret the lack of large \lya EW, 
UV bright galaxies as an observational effect of small survey volumes. 

For the model presented in this paper, the UV luminosity is directly related to
halo mass and the apparent \lya luminosity is determined by radiative transfer,
which depends on the environment of 
galaxies (\S~\ref{sec:factors}). It is interesting to see to what extent the
observed relation between \lya EW and UV luminosity can be explained by our
model. Following \S~\ref{sec:LFlya} and \S~\ref{sec:uvLF}, we scale the
\lya luminosity by a factor of 5 and apply an effective UV extinction of 0.3 
for the UV luminosity for each LAE source.

In our model, the intrinsic \lya EW is a constant for all sources, since 
both the intrinsic \lya luminosity and the UV luminosity are proportional
to the SFR. Radiative transfer, however, gives rise to a broad distribution 
of the apparent (observed) \lya EW. The distribution of the apparent EW 
is similar to that in Figure~\ref{fig:LL}$b$, if the horizontal axis is 
relabeled. Figure~\ref{fig:LL}$b$ shows the distribution in logarithmic 
space, while EW distribution in linear space matches more closely with 
what can be inferred from observations. In Figure~\ref{fig:EW_PDF}, we show
the distribution of rest-frame \lya EW in linear space from our model. At a 
given UV luminosity, the distribution function of apparent \lya EW is a 
decreasing function of EW. In the UV luminosity range considered here, the 
distribution function drops faster for sources with lower UV luminosity.

In the left panel of Figure~\ref{fig:EW_conqua}, the cyan contours show the
probability density distribution of objects in the plane of \lya EW and UV
luminosity from the model (thicker contours for higher densities). The apparent
\lya EW distribution at a fixed UV luminosity roughly follows an exponential 
distribution from our model, with fewer sources having larger EW. If the EW 
distribution did not vary with UV luminosity, the contour of equal probability
density in the left panel of Figure~\ref{fig:EW_conqua} would appear to tilt
along the direction of low EW and UV bright to high EW and UV faint because
the number density of objects drops fast with UV luminosity. In our model, 
the apparent EW distribution has a weak dependence on UV luminosity, but it
only leads to a slight change in tilt direction of the contour and cannot 
mask the effect caused by the decreasing number density toward high UV 
luminosity. The cyan contours in the left panel of Figure~\ref{fig:EW_conqua}
clearly show that sources at the corner of large apparent EW and high UV 
luminosity have a low probability density. The low probability is a 
consequence of the combination of two facts: that UV LF drops steeply towards 
high luminosity and that the distribution of the apparent \lya EW at fixed 
UV luminosity is a decreasing function of EW.
The result suggests that a large survey volume 
is needed to discover large EW, UV bright sources, which is consistent with 
the conclusion in \citet{Nilsson09}. We note that we do not assume any 
particular form of the EW distribution at a given UV luminosity. In our model,
we have a single value of intrinsic EW for all sources, since both intrinsic 
\lya luminosity and UV luminosity are proportional to SFR and we apply the
same scaling in either luminosity for all sources. The distribution of the
apparent EW simply results from the radiative transfer effect.

\begin{figure}
\plotone{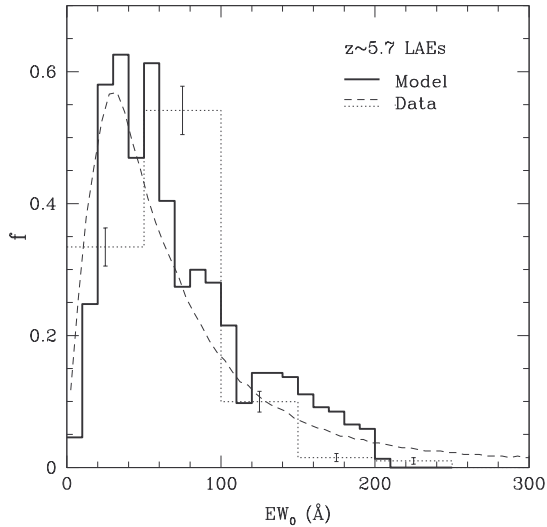}
\caption[]{
\label{fig:EW_pdf}
Distribution of \lya rest-frame equivalent width (EW$_0$) for $z\sim 5.7$ LAEs.
The dashed curve and the dotted histogram are two estimates of the distribution
for all the photometrically selected $z\sim 5.7$ LAEs in \citet{Ouchi08}. 
The dashed curve is estimated from a maximum likelihood method by fully 
accounting for the probability distribution of the measured EW$_0$ for each 
LAE and the dotted histogram is obtained by simply adopting the best 
measurement of EW$_0$ (see \citealt{Ouchi08}). The solid histogram is from
our model for the corresponding population of LAEs, i.e., for LAEs that have 
the same number density as in \citet{Ouchi08}.
}
\end{figure}

Because of the flux limit, luminosities of observed LAEs are above a 
threshold. The red solid curve in either panel of Figure~\ref{fig:EW_conqua}
is the corresponding threshold for EW as a function of UV luminosity
for $z=5.7$ LAEs in SXDS. 
To make a further comparison between the model and the observed LAEs, we show 
in the right panel of Figure~\ref{fig:EW_conqua} the median and quartiles 
(solid and dotted cyan curves) of the EW distribution as a function of UV 
luminosity from the model, for LAEs corresponding to those observed in the 
Subaru fields (filled circles and squares). The model prediction largely 
follows the observational trend. Again the lack of large EW, UV bright LAEs
is evident in the model. Although the small number of observed LAEs prevents
a comparison of the model and observed EW distribution as a function of UV 
luminosity, we can make a comparison for the overall EW distribution of the 
observed LAEs. The dashed curve and dotted histogram in Figure~\ref{fig:EW_pdf}
are two estimates of the EW distribution for all the LAEs detected in SXDS 
inferred by \citet{Ouchi08}. The dashed curve is obtained with a maximum
likelihood method by accounting for the full probability distribution of the 
measured EW for each LAE. The dotted histogram is obtained by simply counting
the number of LAEs in each EW bin based on the measured values of EW, i.e., no 
uncertainty in the measured EW is assumed. According to \citet{Ouchi08}, the 
two estimates likely bracket the true distribution. The corresponding EW 
distribution from our model is shown as the solid histogram, which appears to
be in good agreement with the observation estimates. Interestingly, it is 
more closely resemble the one from the maximum likelihood method.

Without appealing to differences in UV faint and UV bright sources, such as
the amount of dust and age of stellar population, the observed deficit of 
bright UV galaxies with large \lya EW is reproduced by the model. It is a 
natural consequence of the fact that UV LF drops toward high luminosity and
the distribution of apparent EW at fixed UV luminosity is largely a decreasing
function of EW. The model also reproduces the observed EW distribution. We
emphasize that at fixed UV luminosity, the distribution of apparent EW in 
the model is completely determined by \lya radiative transfer effect, not 
by any other mechanisms (e.g., different UV extinction or stellar age). Our 
model suggests that dependence of dust or stellar age on halo mass (or UV 
luminosity), if there is any, does not play a dominant role in the observed
\lya EW distribution of high-redshift galaxies and that \lya radiative 
transfer is the main mechanism in determining the observational properties.

\section{Important Physical Factors in Determining the Observability of LAEs}
\label{sec:factors}

\begin{figure*}
\plotone{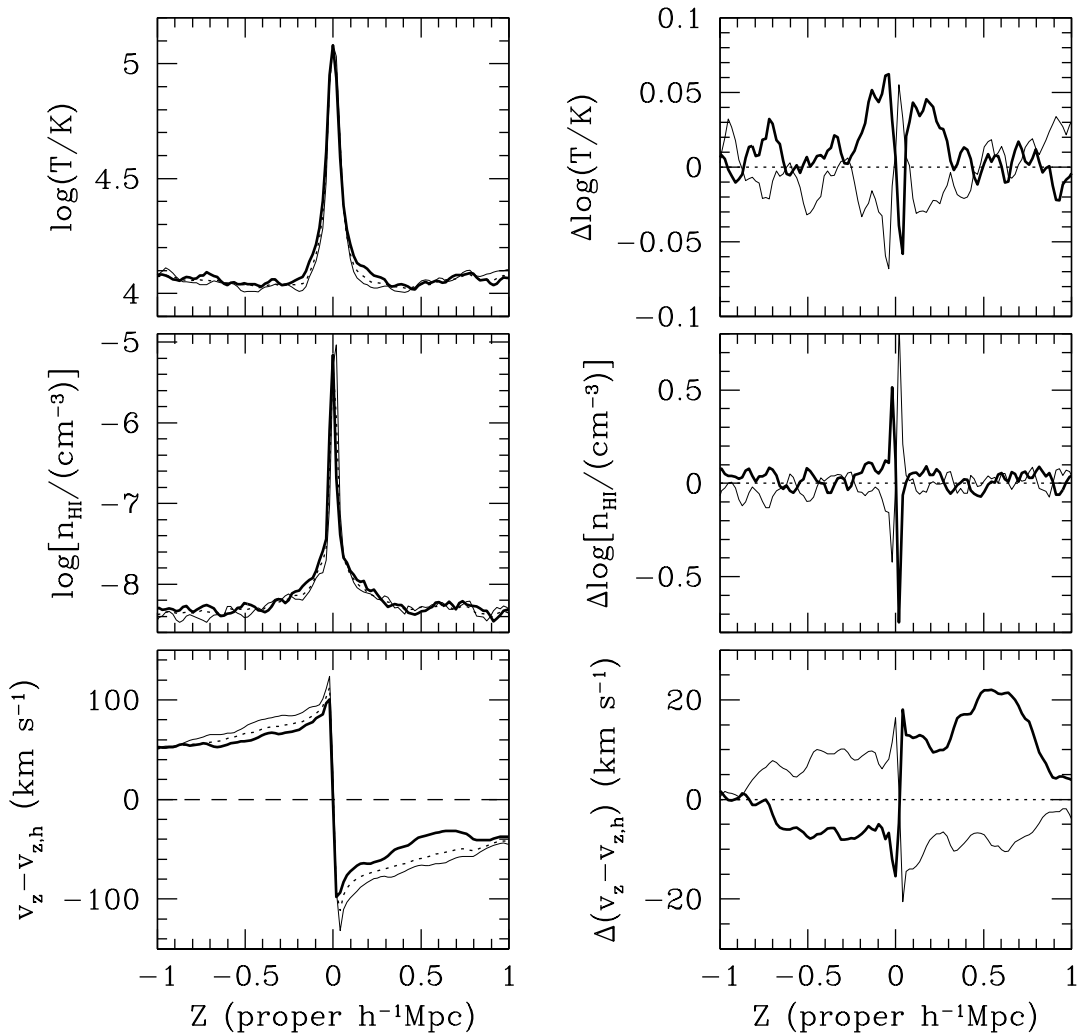}
\caption[]{
\label{fig:halo_profile}
Mean gas distribution profiles along the line of sight around halos of 
mass $8\times 10^{10}\hMsun$. In the left panels, from top to bottom, 
are temperature, neutral hydrogen density, and peculiar velocity profiles.
Note that the halo velocity is subtracted from the peculiar velocity. The 
dotted curve in each panel is the mean profile for all the halos in the 
narrow mass bin around $8\times 10^{10}\hMsun$. The thin and thick solid 
curves are those in the lower and upper quartiles of the apparent to intrinsic
\lya luminosity ratio. In the corresponding panel on the right, we show the
profile with respect to the mean profile, i.e., the mean profile for all 
halos in the mass bin is subtracted. 
}
\end{figure*}

\begin{figure*}
\plotone{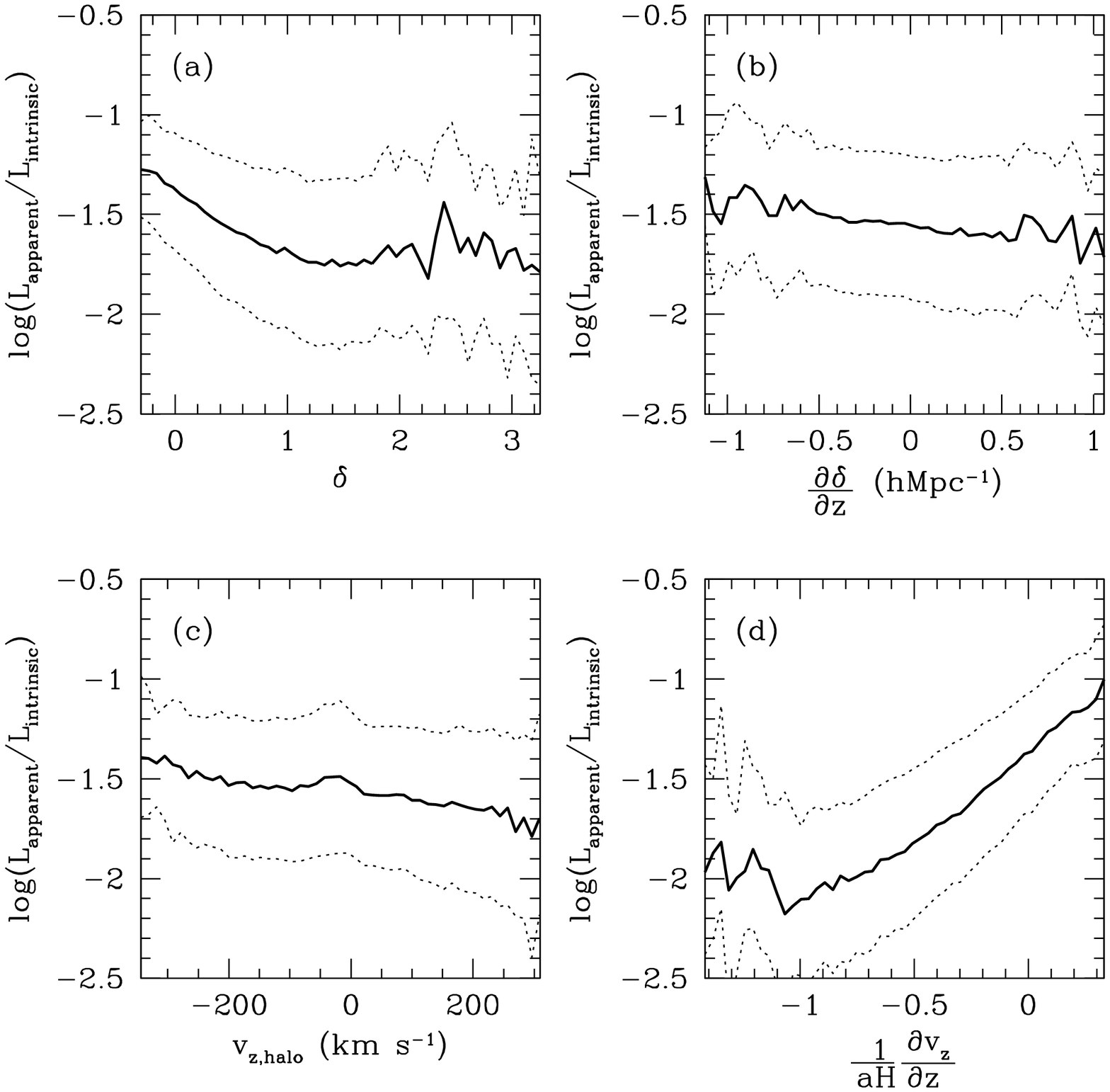}
\caption[]{
\label{fig:denvelgrad}
Dependence of \lya flux suppression of LAEs on density and peculiar velocity.
The suppression is characterized by the ratio of the apparent (observed) and 
intrinsic \lya luminosity $L_{\rm apparent}/L_{\rm intrinsic}$. 
{\it Panel~(a)}: dependence on the smoothed overdensity field at the source 
position. The overdensity field is smoothed with a 3D top-hat filter of radius
2$\hMpc$ (comoving), which is chosen to be larger than the size of the infall
region around halos.
{\it Panel~(b)}: dependence on the density gradient along the $Z$ direction. 
The derivative is with respect to comoving coordinate. 
{\it Panel~(c)}: dependence on the host halo velocity. 
{\it Panel~(d)}: dependence on the linear peculiar velocity gradient along the 
$Z$--direction. The linear peculiar velocity is obtained from the smoothed 
overdensity field based on the continuity equation (see text for detail). 
The velocity gradient is put in units of the Hubble parameter.
In each panel, the solid curve is the median ratio and the two dotted 
curves delineate the upper and the lower quartiles. Note that the 
observation is along the $-Z$ direction, which matters for interpreting
the results in of panels~($b$) and ($c$). See the text.
}
\end{figure*}

In previous sections, we study statistical properties of \lya emission
of LAEs in our model. At a fixed intrinsic \lya luminosity, the apparent
luminosity and peak wavelength shift have broad distributions. The 
cause of the distribution must be related to the underlying distribution and
kinematics of gas, hence the matter distribution, around LAE sources, 
since \lya radiative transfer is sensitive to the density and velocity fields.
In this section, we study the correlations between the apparent \lya 
properties and the environment of matter around sources. By revealing the
physical causes, such correlations would aid our understanding of the 
observability and the statistical properties of LAEs.

To identify the key factors in shaping the observability of LAEs, we
first choose LAE host halos in a narrow mass bin 
($\sim 8\times 10^{10}\hMsun$) and stack the neutral gas density, temperature,
and peculiar velocity profiles along the line of sight centered on these halos. 
As shown by the dotted curve in the middle-left panel of 
Figure~\ref{fig:halo_profile}, the stacked density profile around 
a halo appears to peak at the halo center and to be symmetric around the 
center. The central density is about a factor of 2000 higher 
than the mean cosmic density. We then stack only sources that are 
strongly and weakly suppressed in \lya luminosity (e.g., the upper and lower 
quartiles of $L_{\rm apparent}/L_{\rm intrinsic}$), respectively, into two 
subsets. The stacked density profiles of strongly and weakly suppressed 
sources (thin and thick solid curves in the middle-left panel of 
Fig.~\ref{fig:halo_profile}) appear to have a relative offset in amplitude 
and are asymmetric in opposite directions. The trend becomes clear with
the total mean profile subtracted (middle-right panel of 
Fig.~\ref{fig:halo_profile}). 
The temperature profiles of strongly and weakly suppressed \lya sources
also show systematic but small differences (top panels of 
Fig.~\ref{fig:halo_profile}). As temperature is largely linked to density 
inside halos, we tend to incorporate its effect into an overall density 
effect. The difference in amplitude and asymmetry of density profiles between
strongly and weakly suppressed sources indicates that density and density 
gradient along the line of sight contribute to the observability of LAEs.

The overall stacked peculiar velocity profile around a halo (bottom-left
panel of Fig.~\ref{fig:halo_profile}), with the
velocity centered on the halo velocity, shows clear
signatures of infall region: an inner Hubble-like contraction region near 
the center and an outer region with infall velocity decreasing outward.
Similar to the density profile, we find difference in the slope of the 
velocity profile in the outer infall region between strongly and weakly 
suppressed \lya sources (bottom-right panel of Fig.~\ref{fig:halo_profile}). 
Therefore, the peculiar velocity gradient along the line of sight is a factor 
in determining the observability of LAEs. We check the dependence on the 
line-of-sight velocity of halos and find that halo velocity also contributes 
to the LAE observability. 

The above exercises provide us with the initial evidence
on what physical variables affect the observability of LAEs. 
Since on scales larger than halo size, gas density 
and velocity largely follow those of the underlying dark matter, we
identify the matter density and peculiar velocity and their gradients along the
line of sight as the major factors in shaping the LAE observability.
We proceed to study the correlation between these quantities and the
suppression in \lya luminosity for all the sources in our model. We 
are interested in the environment density and velocity around sources, not 
those inside halos, but they cannot be obtained directly at the positions of 
halos from the outputs of the reionization simulation. To eliminate the 
influence of density and velocity profiles inside halos, we smooth the density 
field with a 3D top-hat filter of radius of 2$\hMpc$ (comoving). 
Conclusions reached below are largely immune to possible uncertainties 
related to limited resolution of the hydro simulations we use 
($\sim$65kpc comoving), which nonetheless is much smaller than 2$\hMpc$ 
comoving.

With the smoothed overdensity field $\delta$, we solve the linear peculiar 
velocity from the continuity equation
\begin{equation}
\label{eqn:continuity}
\dot\delta+\frac{1}{a}\nabla\cdot{\mathbf v}=0,
\end{equation}
where $\dot\delta$ can be written as $fH(a)\delta$ with $f=d\ln\delta /d\ln a$
the derivative of the growth factor and $H(a)$ the Hubble parameter at the 
time when the scale factor is $a$. The peculiar velocity field is obtained
from the Fourier transform
\begin{equation}
{\mathbf v}=fHa \sum_{\mathbf k}\frac{ik_z}{k^2}\delta_{\mathbf k} 
e^{i{\mathbf k}\cdot{\mathbf r}},
\end{equation}
where $\delta_{\mathbf k}$ is the Fourier transform of the smoothed overdensity
field. The velocity gradient and density gradient along the $Z$ direction are
\begin{equation}
\frac{\partial v_z}{\partial z}=
-fHa\sum_{\mathbf k}\frac{k_z^2}{k^2}\delta_{\mathbf k} 
e^{i{\mathbf k}{\mathbf \cdot}{\mathbf r}}
\end{equation}
and
\begin{equation}
\frac{\partial \delta}{\partial z}=
\sum_{\mathbf k}i k_z\delta_{\mathbf k} 
e^{i{\mathbf k}\cdot{\mathbf r}}.
\end{equation}
The spatial derivatives in all the above equations are with respect to comoving
coordinates.

Figure~\ref{fig:denvelgrad} shows the correlations of the apparent to intrinsic
luminosity ratio with the environment overdensity, velocity and their gradient
along the line of sight for all the LAEs in our model. The trends of 
correlations at a fixed halo mass look similar, with amplitudes and slopes 
of curves slowly evolving with mass. These correlations reflect different 
aspects of \lya radiative transfer. In what follows, we interpret them in turn.

Figure~\ref{fig:denvelgrad}$a$ shows that 
\lya emission from LAEs is more strongly suppressed in denser region.
This seems easy to understand --- higher 
density means high optical depth for \lya emission. 
At the high end of
the density, high optical depth also leads to large \lya frequency diffusion
(and more spatial diffusion),
which appears to be able to compensate for otherwise reduced \lya transmission 
due to high optical depth. 
As a result, the curve of the overall luminosity suppression factor 
flattens at very high density.
In details, the above understanding is not complete. Through the continuity
equation (eq.[\ref{eqn:continuity}]), density is anti-correlated with velocity 
gradient. As shown below (Fig.~\ref{fig:denvelgrad}$d$), velocity gradient 
plays a dominant role in determining the suppression factor.
As will be elaborated in Paper II, the density dependence seen here is 
largely driven by its anti-correlation with the velocity gradients, while 
velocity gradients along different directions contributes a lot to the 
anisotropic \lya emission distribution.

Figure~\ref{fig:denvelgrad}$b$ shows that
\lya sources are easier to be transmitted if the local density has a negative 
gradient along the line of sight, i.e., density decreases towards the observer. 
Note that the observation direction (line of sight) for \lya emission in our radiative
transfer modeling is set to be along the $-Z$ direction, therefore negative
gradient means that sources are located on the near side of overdense
regions or on the far side of underdense regions. 
With respect to sources located on the far side of overdense
regions or on the near side of underdense regions,
\lya photons encounter a decreasing density
profile, thus low optical depth, along their way to the observer. Although 
these photons also travel across other overdense and underdense regions,
the redshifting of these photons caused by Hubble expansion would make these
intervening regions largely transparent to them. That is, only the immediate 
environments of the sources affect their observability.

Figure~\ref{fig:denvelgrad}$c$ show that, 
with respect to the Hubble flow, LAEs moving away from the observer have a
lower suppression in \lya luminosity than those moving towards the observer.
Note that sources with negative velocities 
are the ones moving away from the observer, since we observe along the $-Z$
direction. The IGM on scales larger than LAE host halos can detach from 
the motion of halos to some degree, and experiencing Hubble expansion. 
The motion of the halo with respect to the surrounding IGM thus introduces
a dipole in the \lya optical depth around the source. For sources moving 
away from the observer, \lya photons acquire an additional redshift from the 
halo motion, and hence the direction towards the observer is the one that 
has the lower optical depth and photons preferentially ``leak" towards 
that direction.

Figure~\ref{fig:denvelgrad}$d$ shows that the line-of-sight gradient 
of the line-of-sight velocity has the largest effect in shaping the LAE observability. 
We see that sources located at places that have larger line-of-sight gradient in peculiar
velocity are easier to be observed. 
A local velocity gradient effectively
changes the local Hubble expansion rate. A positive gradient increases the
local expansion rate. A faster expansion makes \lya photons on the red side 
of line center much easier to escape. It also makes \lya photons on the blue
side of line center travel a shorter distance to redshift to the
line center and to be scattered in the IGM, and therefore the escaping photons 
after frequency diffusion are more centrally distributed, leading to 
higher surface brightness. 
Both of these effects cause the transmission of \lya photons that is 
enhanced for sources with positive local velocity gradient. 

We see that the major factors in determining LAE observability all have
clear physical origins. Quantities like the density gradient, velocity, 
and velocity gradient are statistically interrelated. For example,
in a statistical sense, sources on the near side of an overdense region
(negative density gradient) usually moving away from us (negative velocity).
However, on a source by source basis, because of the randomness of the
density and velocity field, this is not always true. In other words, there are
large scatters among the correlations of these quantities. Since these different
quantities have different physical effects on \lya transmission, the overall
\lya transmission or luminosity suppression effect should be a supposition of 
all of them. 

The dependence of \lya radiative transfer on large scale density and peculiar 
velocity fields imposes a strong selection effect on observations of LAEs.
The selection leads to new features in the clustering of LAEs, which are
investigated in Paper II.

Some of the environment factors identified here are expected to be found in 
the $\exp(-\tau_\nu)$ model. For example, \citet{Iliev08} compare the results 
from the $\exp(-\tau_\nu)$ model with and without the peculiar velocity field 
turning off. They find that both the gas infall around halos and source 
peculiar velocity are important in determining the suppression factors of 
\lya luminosity and in shaping the observed \lya luminosity, especially at the
luminous end (see their Figures 21, 23 and 24). This is qualitatively in 
parallel with our finding that source peculiar velocity and peculiar velocity 
gradient of surrounding matter are important factors. While \citet{McQuinn07} 
consider the cases of varying the global neutral fraction, which shows the 
effect of density on the suppression of \lya luminosity, it is clearly 
different from what we intend to do here. Nevertheless, we expect that the 
density and density gradient effects also show up in the $\exp(-\tau_\nu)$
model, given that they contribute to the line-of-sight optical depth.
Although in most cases the simple $\exp(-\tau_\nu)$ model can provide 
qualitative understanding of \lya radiative transfer results, we do not 
expect it to capture all the physics. In the $\exp(-\tau_\nu)$ model, the 
radiative transfer is completely determined by the {\it line-of-sight}
optical depth. In the full calculation, this is not the case -- the scatterings
of \lya photons enable them to probe the optical depth in all directions
and the {\it line-of-sight} outcome depends on not only the {\it line-of-sight}
optical depth but also those in other directions (see more details in Paper II).

\section{Summary and Discussion}

\subsection{Summary of Main Results}

We perform a full \lya radiative transfer calculation with a Monte Carlo code
\citep{Zheng02} to study LAEs in a cosmological volume. The LAE sources and 
the physical properties of neutral hydrogen gas are taken from the $z\sim 5.7$ 
outputs of a cosmological reionization simulation \citep{Trac08}, which solves 
the coupled evolution of the dark matter, baryons, and ionizing radiation in a
box of 100$\hMpc$ (comoving) on a side. The large volume of the simulation 
allows a statistical study of $z\sim 5.7$ LAEs. Radiative transfer of \lya 
photons in the IGM environment around LAEs, which leads to both frequency and 
spatial diffusion of \lya photons, turns out to play a crucial role in 
determining the observability of LAEs and in understanding the observed 
properties of LAEs.

Although the radiative transfer calculation is computationally costly, 
the LAE model we 
present in this paper is rather simple. The UV or intrinsic \lya luminosity 
is assumed to be proportional to the SFR, which is tightly coupled to halo 
mass in the reionization simulation we use. That is, we essentially adopt a 
constant mass to light ratio, where mass is halo mass and light is either UV 
or \lya. All we do is to add the physics of \lya radiative transfer into the 
model to obtain the observed properties of LAEs. That is, we introduce the 
radiative transfer of \lya photons in the IGM as the single factor responsible 
for transforming the intrinsic \lya emission properties to the observed ones. 
Our model produces IFU-like data cube that covers the extent of the simulation 
box, which allows mock observations to be made. With the \lya image contracted 
from this data cube, we follow typical observational procedures 
\citep[e.g.,][]{Ouchi08} to identify LAEs and then extract their \lya spectra.

Initially \lya photons are produced inside the star formation region. Therefore
the intrinsic \lya sources are expected to be similar in size as the UV 
sources, which are compact ($\lesssim 1$ kpc; \citealt{Taniguchi09}). We find 
that an intrinsically point-like \lya source becomes extended as a consequence 
of resonant scatterings of \lya photons (spatial diffusion). The scatterings
of \lya photons do not destroy them and all \lya photons escape in the end.
However, observationally, only the central part of the extended source can be 
detected as a consequence of the limit set by the surface brightness threshold.
The scatterings of \lya photons also cause the frequency of \lya photons to 
change (frequency diffusion). The resultant \lya spectra from the central 
aperture do not have a simple relation to the initial profile, which is 
assumed to be Gaussian in our model. Our results from full radiative transfer 
calculations show a clear difference from a simple treatment of 
\lya radiative transfer, namely the $\exp(-\tau_\nu)$ model, widely adopted 
in previous work, which modifies the intrinsic \lya spectrum by multiplying 
the line-of-sight transmission determined by the optical depth at each 
frequency.

The observed \lya spectrum of an LAE in our model shows a clear asymmetry, 
skewed towards red. Although the $\exp(-\tau_\nu)$ model produces the same 
qualitative feature, the predicted line profile, the frequency shift, and 
the total flux are all significantly different from our results. 
While the spectrum of the $\exp(-\tau_\nu)$ model we present is essentially 
the intrinsic one truncated below a certain wavelength (but see 
Figures 14 and 15 in \citealt{Iliev08} for more complex line shapes, 
probably caused by different assumptions in the $\exp(-\tau_\nu)$ model), 
the observed \lya spectrum in our model can have
contributions from photons with frequency much redder than initial photons,
a result of the scattering-caused frequency diffusion. 
We find that the redward shift of the Lya line induced by radiative transfer 
is usually a few times the intrinsic line width, with a distribution that 
peaks at about three times. The asymmetry and shift of the \lya line do not 
indicate the presence of any winds, but they arise from the structure of 
the halo infall and Hubble expansion around the sources. 
If one were to infer the wind velocity, if there is any,
from comparing the relative shift in the \lya line and an optically-thin line,
one has to keep in mind the \lya radiative transfer effect. For example,
the observationally inferred velocity of the receding winds would be 
overestimated by $\sim 100\kms$ or more if the effect is not taken into 
account.

As a consequence of the frequency diffusion and spatial diffusion, our model 
predicts a much higher observed \lya flux than the $\exp(-\tau_\nu)$ model. 
At a fixed intrinsic \lya luminosity (i.e., fixed host halo mass), the 
observed (apparent) luminosity is broadly distributed. The shift in the
\lya line peak and the ratio of the apparent to intrinsic \lya luminosity
appear to be anti-correlated. The distributions of the line peak shift and 
the ratio of apparent to intrinsic \lya luminosity, and their correlation,
all result from the dependence of the \lya radiative transfer on the 
IGM environments around sources. It would be interesting and extremely useful 
if we could make use of the full information in the observed \lya properties 
to infer the intrinsic ones, and we reserve such an investigation for future
work. Although our model predicts a much higher observed \lya flux than the 
$\exp(-\tau_\nu)$ model, it still leads to a highly suppressed \lya flux,
compared with the intrinsic one. The suppression factor depends on the 
assumed line width of the intrinsic \lya spectra. For the line width assumed
in our model (given by halo virial temperature), we find that, with respect 
to the intrinsic \lya LF of LAEs, the observed (apparent) \lya LF shift 
towards the low luminosity end by roughly one order of magnitude in 
luminosity. For comparison, the $\exp(-\tau_\nu)$ model would shift by two 
orders of magnitude in luminosity.

We make comparisons between the $z\sim 5.7$ LAEs in our model and those 
observed in SXDS \citep{Ouchi08}. The sizes, morphologies, \lya line profiles 
of the model LAEs are remarkably similar to the observed ones. 
For the \lya LF, UV LF, and
\lya EW distribution, our model can successfully reproduce the observations 
and provide physical explanations for various observed features. 

After an overall adjustment of a factor of $\sim 5$ in luminosity, the \lya 
LF of model LAEs matches well with observation. The adjustment reflects our 
incomplete knowledge in the stellar IMF at high redshift, the uncertainty in 
the model SFR, and the lack of information on the intrinsic \lya line profile. 
According to our model, there is no one-to-one map between the intrinsic and 
the observed \lya luminosity. In other words, there is a large scatter in the 
relation between the apparent and intrinsic luminosities. At a fixed observed 
luminosity, LAEs can differ by one order of magnitude in the intrinsic 
luminosity (Fig.~\ref{fig:LL}$c$). This large scatter has to be taken into 
account when interpreting the observed \lya LF and linking the observed LAEs 
to their host halos.

For the UV LF of observed LAEs, our model prediction shows a good agreement 
with observation. In particular, the turnover of the UV LF towards the low 
luminosity end seen in high-$z$ ($z\sim$ 3--6) LAEs are well reproduced.
The key to interpret the shape of the UV LF is that observed LAEs are 
sources with observed (apparent) \lya luminosity above certain threshold.
The turnover reflects that for LAEs with low UV luminosity (or low 
intrinsic \lya luminosity, or low halo mass), the probability for the
observed \lya luminosity to exceed the observation threshold
is low, a consequence of the broad distribution of apparent \lya luminosity
at a given intrinsic \lya luminosity. The full UV LF for sources in our 
model (i.e., without imposing the observation \lya luminosity threshold) 
agrees well with the nonduplicated sum of the observed UV LFs of LAEs and 
$i$--dropout galaxies at $z\sim 6$.

The observed distribution of \lya EW as a function of UV luminosity is also
reproduced in the model. We note that in our model all the sources have the 
same {\it intrinsic} EW and the distribution of the {\it observed} values 
of EW at fixed UV luminosity is purely caused by the environment-dependent 
radiative transfer effect. At a fixed UV luminosity, the distribution of 
observed (apparent) EW is a decreasing function toward high values.
The observational trend of lacking UV bright, high EW sources 
\citep[e.g.,][]{Ando06} is naturally explained by our model in that
such sources lie in a low probability corner --- a combination of the
drop of the UV LF toward high luminosity and the drop of the apparent EW 
distribution function toward high EW value. LAE surveys with large volume
will test the interpretation.

Therefore, the observed properties of LAEs can be explained by simply invoking 
\lya radiative transfer: the effects of the local IGM environment, depending 
mainly on the gas density and line-of-sight velocity and their line-of-sight 
gradients, lead to the distribution of observed \lya emission properties at 
fixed intrinsic \lya luminosity. This environmental selection also causes 
new features in the clustering of LAEs that we will study in Paper II.

\subsection{Implications and Discussion}

Our interpretation of the observations of LAEs does not invoke any mass 
dependent dust absorption, which is in contrast to many previous models 
\citep[e.g.,][]{Dayal10}. Uniformly distributed dust efficiently absorbs 
\lya photons, since the large number of resonant scatterings increase the
path length. There is much less attenuation when the dust is in gas clumps 
and \lya photons bounce off the cloud surfaces \citep{Neufeld91,Hansen06}.
Optical, UV, and \lya observations of local star-forming galaxies provide 
evidence that ISM kinematics and geometry play a more significant role than 
dust in affecting the \lya emission \citep[e.g.,][]{Giavalisco96,Keel05,Atek08,
Atek09}. Our model successfully reproduces the observed UV LF of LAEs 
by incorporating only a mass independent effective UV extinction of at 
most 0.3 mag. We conclude that any mass dependent dust effects are not 
likely to play a substantial role to determine the observed properties of 
LAEs, compared to \lya radiative transfer effects.

Our model also has important implications for the duty cycle and the \lya 
escape fraction of LAEs. The theoretically predicted (intrinsic) \lya LF, 
which essentially is the halo mass function, is substantially higher 
than the observed one. Two scenarios have been introduced to address this
problem, the duty cycle and the \lya escape fraction scenarios 
\citep[e.g.,][]{Stark07b,Nagamine08}. In the duty cycle scenario, LAEs are 
short-lived and a fraction of all galaxies are active as LAEs at any given 
time, lowering the amplitude of the \lya LF. In the \lya escape fraction 
scenario, only a fixed fraction of \lya photons escape from the source
and the overprediction problem is solved by shifting the LF towards the 
low luminosity end. To conserve the number density of LAEs of a given sample, 
the masses of host halos in the duty cycle scenario would be on average lower 
than those in the escape fraction scenario. As a consequence, the clustering 
of LAEs would be different in the two scenarios, with a stronger clustering in 
the escape fraction scenario. \citet{Nagamine08} find that LAE clustering 
measurements from observations are in favor of their duty cycle scenario.

In our model, \lya photons all escape after a large number of scatterings. The
\lya escape fraction, in its literal meaning, is therefore unity. However, 
only the central part of the extended \lya emission of LAEs can be observed,
which gives rise to an {\it apparent} or {\it effective} \lya escape fraction.
Since the observed \lya luminosity has a broad distribution at a fixed 
intrinsic \lya luminosity, our model predicts a broad distribution of the 
effective \lya escape fraction rather than a single value. In our model, no 
duty cycle parameter is introduced. Since halos of the same mass have similar
SFR in our model, the corresponding intrinsic \lya luminosities are the same,
i.e., \lya emission does not come from a fraction of halos. However, an 
{\it apparent} or {\it effective} duty cycle arises as a result of the 
selection effect caused by \lya radiative transfer (a broad distribution of 
observed \lya luminosity at a fixed intrinsic \lya luminosity) and a \lya 
luminosity threshold in observation. This can be seen from comparing the 
UV LF for all galaxies (dropout galaxies and LAEs) and that for LAEs
(Fig.~\ref{fig:uvLF}), which can be described as that at a fixed UV luminosity
(or halo mass) only a fraction of all the galaxies are observed as LAEs.
This effective duty cycle does not have the physical meaning in its original 
form. Moreover, it is not a constant, since it changes with UV luminosity 
(Fig.~\ref{fig:uvLF}). \citet{Tilvi09} present an LAE model in which \lya 
luminosity (or SFR) is related to the halo mass accretion rate, rather than 
halo mass, and the model naturally gives rise to the duty cycle of LAEs. 
The duty cycle in their model, however, has its original meaning, in direct 
contrast with our model. Our model still ties the intrinsic \lya luminosity 
(SFR) to halo mass and let the \lya radiative transfer do the work of 
converting it to observed \lya luminosity. Because of the large scatter 
between the observed \lya luminosity and the intrinsic one (or halo mass)
in our model, for a sample of LAEs above a \lya luminosity threshold, some 
of them can reside in halos with mass smaller than the threshold mass above 
which halos have the same number density as LAEs. So the effective duty cycle 
in our model has the effect of lowering the clustering amplitude of LAEs. 

The main uncertainties in our model are the stellar IMF, the SFR, and the 
intrinsic \lya line profile. The first two are general uncertainties for
any model. The IMF at high-$z$ is neither well constrained observationally 
nor well understood theoretically. The SFR in galaxy formation model is 
related to the complex gas physics that we do not have a satisfactory 
understanding. Changing IMF or SFR would change the details in the 
reionization process and therefore change the gas properties (e.g., neutral 
fraction and temperature distribution) at a fixed redshift. 
For the reionization history itself, the escape fraction of ionizing photons
adds a further uncertainty.
Even though we focus on LAEs at $z\sim 5.7$, when reionization is almost 
complete, different reionization histories can still leave different imprints
on the gas distribution. For example, the IGM temperature in a region is 
correlated to the time when this region is reionized and heated 
\citep[e.g.,][]{Trac08}. A detailed study is needed to investigate the effect 
of inhomogeneous IGM temperature distribution on \lya radiative transfer.   
To be fully self-consistent, for any change in the IMF, SFR, and escape 
fraction of ionizing photons, one has to re-run the reionization simulation
to solve the density, velocity, and temperature distributions of neutral gas
and then perform the \lya radiative transfer calculation.
If the IMF, SFR, and escape fraction of ionizing photons change in a way to 
maintain the same reionization history, the effect of 
the IMF and SFR change can be largely characterized by an overall scaling in 
UV or intrinsic \lya luminosity and one does not need to redo the 
\lya radiative transfer calculation. For simplicity and to avoid 
extensive computations for reionization and radiative transfer simulation, 
we adopt such a scenario in this paper. 
Changing the width of the intrinsic \lya line profile leads to 
changes in the distribution of the apparent to intrinsic \lya luminosity 
ratio. Although we cast the effect as an overall scaling in the apparent 
\lya luminosity for the \lya LF, the intrinsic \lya line profile is important 
in many aspects of the \lya observation (image, spectra, etc) and its effect 
deserves a detailed investigation. \lya radiative transfer calculation with 
high resolution hydrodynamic simulations for individual LAEs are necessary 
to shed light on the intrinsic \lya line profile \citep[e.g.,][]{Laursen07}. 
Changing the IMF affects the flux of ionizing photons more than          
that of the $\sim$ 1500\AA UV photons, which leads to a change in the ratio 
of the intrinsic \lya luminosity to UV luminosity, or the intrinsic \lya EW.
Intrinsic \lya line profile and dust play a role in converting the intrinsic
EW to apparent EW, with the former affecting the \lya luminosity 
and the latter adding extinction to UV luminosity. Therefore, the full 
distribution of observed \lya EW and UV luminosity of LAEs
can give constraints on the IMF, SFR, dust, and intrinsic \lya profile. 

Although we have identified key environment factors in shaping the 
observational properties of LAEs, the dependence of radiative transfer on 
environments deserves a further study to understand the details of \lya 
scatterings in the surrounding regions of LAEs.

Our \lya radiative transfer calculation relies on the gas distribution
and properties from the cosmological reionization simulation. The radiative
transfer of ionizing photons with a ray tracing algorithm is crucial in
determining the state of gas. We have tested our LAE model for a reionization 
simulation with improved ray tracing algorithm (Trac et al. in prep.) in 
a small box (25$\hMpc$ on a side). We find that the results presented in
this paper are robust.

\lya radiative transfer through the surrounding circumgalactic and 
intergalactic media is a physical process that likely plays an important role 
in galaxies at all redshifts. It has to be taken into account for modeling 
LAEs and for interpreting observations. Our model is rather {\it simple} and 
can naturally explain an array of observations of LAEs, which make it 
extremely attractive. It is interesting to see how well it does in 
interpreting observations of LAEs at lower redshifts (e.g., $z\sim 3$). We 
also plan to apply it to the era of the late stage of reionization to study 
how to use LAEs to constrain reionization.

\acknowledgments

We are grateful to Masami Ouchi for valuable conversations on various 
observational aspects of LAEs, for kindly providing the observational 
data in electronic form, and for helpful comments on an earlier draft. 
We thank Andrei Mesinger for useful discussions and David Weinberg for
helpful comments. 
ZZ thanks Juna Kollmeier and Rashid Sunyaev for useful discussions about 
\lya radiative transfer. 
We thank the referee for useful comments.
ZZ gratefully acknowledges support from the Institute for Advanced Study 
through a John Bahcall Fellowship at an early stage of this work and 
support from Yale Center for Astronomy and Astrophysics through a YCAA 
fellowship. 
ZZ would like to acknowledge the hospitality of the Kavli Institute for 
Astronomy and Astrophysics at Peking University (KIAA-PKU), where part of 
the work was done. 
ZZ also wishes to thank Aspen Center for Physics for a stimulating atmosphere 
and Mark Dijkstra and Paul Shapiro for interesting discussions. 
JM thanks the Institute for Advanced Study for their hospitality. 
This work is supported in part by NASA grant NNG06GI09G. 
HT is supported by an Institute for Theory and Computation Fellowship. 
JM is supported by the International Reintegration Grant of the 
European Research Council 2006-046435 and Spanish grant AYA2009-09745.
This research was supported in part by the National Science Foundation 
through TeraGrid resources provided by NASA. The \lya radiative transfer 
computations were performed at the Princeton Institute for Computational 
Science and Engineering (PICSciE). Some analyses were conducted at the 
aurora cluster at the Institute for Advanced Study.

\appendix
\section{Effects of the Initial \lya Line Width and Gas Temperature in the 
$\exp(-\tau_\nu)$ Model}
\label{sec:expmtau_test}

\begin{figure}
\plottwo{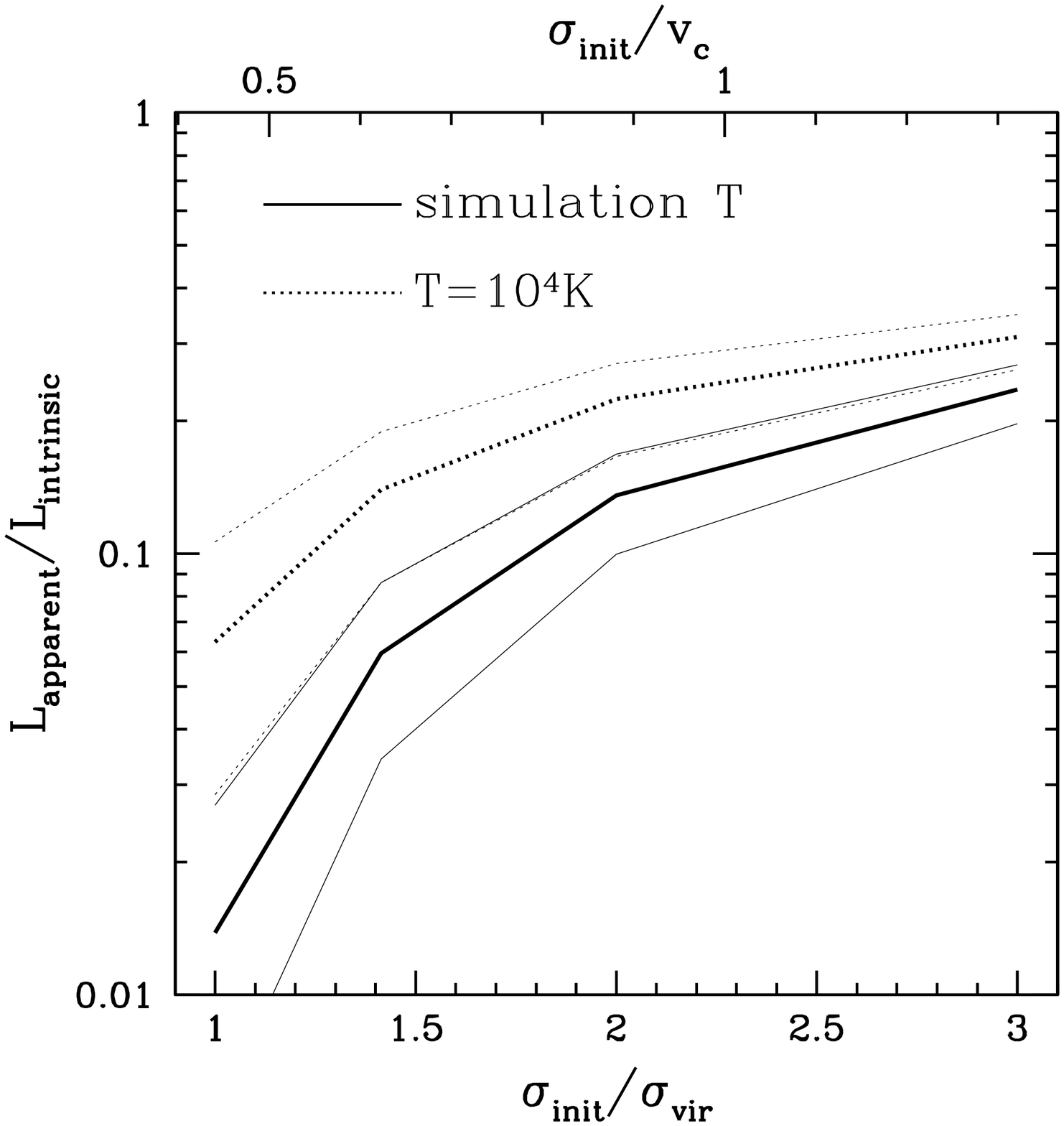}{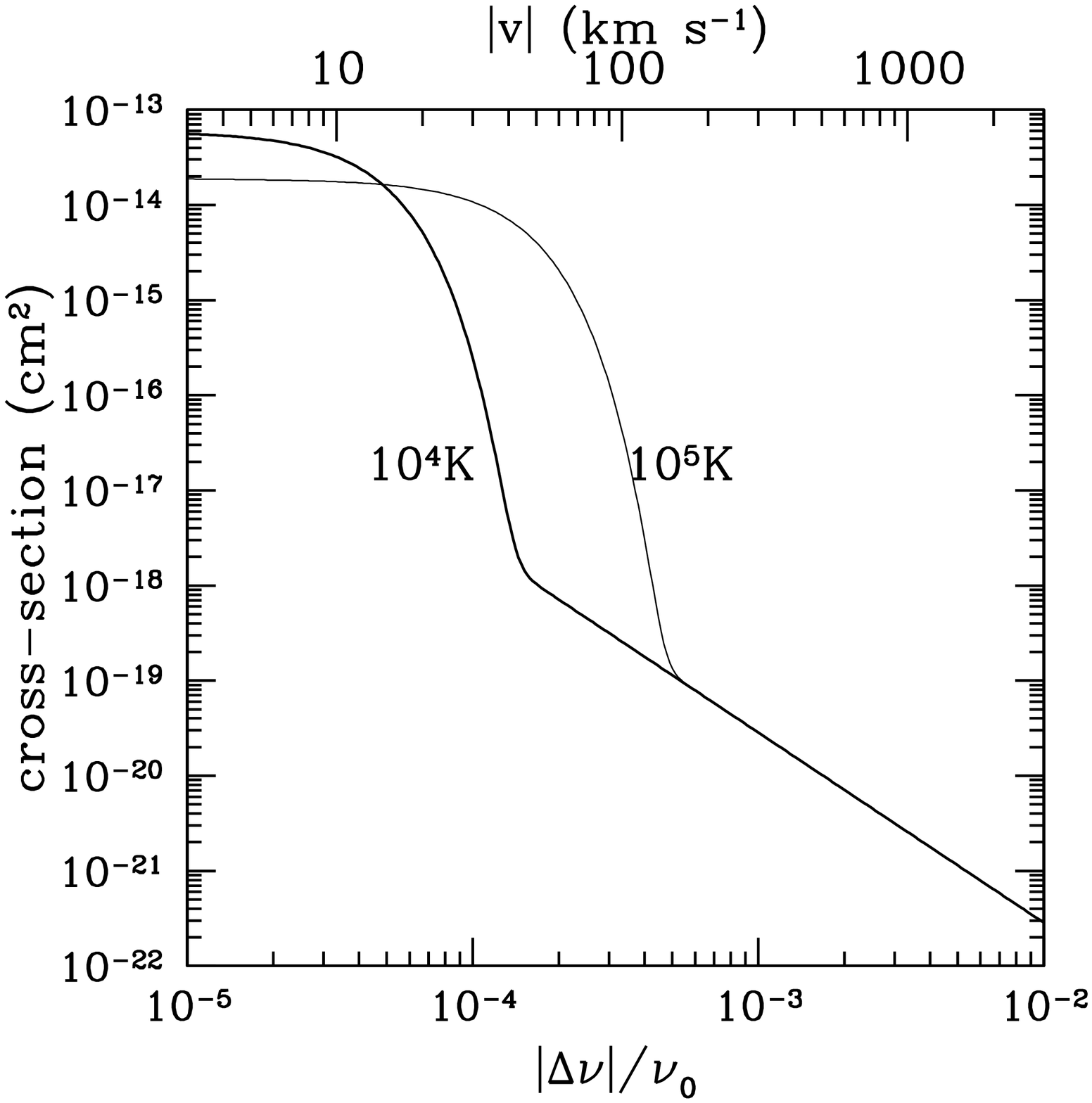}
\caption[]{
\label{fig:expmtau_test}
Effects of the initial \lya line width and gas temperature in the 
$\exp(-\tau_\nu)$ Model.
{\it Left panel:} Ratio of the apparent (observed) to intrinsic \lya luminosity 
as a function of the initial \lya line width $\sigma_{\rm init}$. The bottom
axis marks the line width in terms of the width $\sigma_{\rm vir}$ set
by halo virial temperature, while the top axis in units of the circular
velocity $v_c$ at halo virial radius. The thick and thin solid curves are 
the median and quartiles of the luminosity ratio distribution for the case
using gas temperature in the reionization simulation. The set of dotted
curves are the same but with the gas temperature set to $10^4$K in all regions.
{\it Right panel:} Thermally averaged cross-section of \lya scattering as 
a function of gas temperature. The bottom axis marks the frequency offset 
with respect to the \lya line center and the top axis indicates the fluid
velocity that can give rise to such an offset. The cross-section is composed
of a thermal (Gaussian) core and a Lorentzian wing dominated at small and 
large frequency offsets, respectively. Changing gas temperature from $10^5$K
to $10^4$K has a large effect on the \lya scattering cross-section for gas 
with velocity in the range of $\sim$10--100$\kms$. See text for details.
}
\end{figure}

In \S~\ref{sec:statistics}, we make comparisons of the results from full 
radiative transfer calculation and those from the $\exp(-\tau_\nu)$ model
(see Figures~\ref{fig:LL_cmp}--\ref{fig:LF}). The $\exp(-\tau_\nu)$ model
has been widely adopted in studying LAEs. While the $\exp(-\tau_\nu)$ model 
we present in this paper predicts that the apparent \lya luminosity is at the 
level of a few percent of the intrinsic one, leading to large offset of the 
apparent \lya luminosity function with respect to the one measured from 
observation, the $\exp(-\tau_\nu)$ model in other work 
\citep[e.g.,][]{McQuinn07,Iliev08} does not show such a large suppression in 
\lya flux. For example, the apparent-to-intrinsic \lya luminosity ratio is at 
the level of tens of percent for models in \citet{Iliev08}. The appendix aims
at resolving the discrepancy by clarifying the assumptions adopted in our
$\exp(-\tau_\nu)$ model and others.

In our $\exp(-\tau_\nu)$, we assume the same initial \lya line width as with
the full radiative transfer calculation. The initial rms line width is 
determined by the halo virial temperature (see \S~\ref{sec:RT}),
$\sigma_{\rm init}=31.9[M_h/(10^{10} h^{-1}M_\odot)]^{1/3} \kms$.
We also adopt the gas temperature provided by the reionization simulation, 
which is solved self-consistently in the simulation.
As shown in Figure~\ref{fig:halo_profile}, the temperature of the gas drops
from $\sim10^5$K near the source center to $\sim 10^4$K in the IGM.
For comparison, the initial line width and gas temperature adopted in other
work are different from ours. For example, the fiducial model in 
\citet{McQuinn07} adopt a line width set by the circular velocity at halo 
virial radius, which is about 2.3 times higher than our value. 
\citet{Iliev08} assume the initial rms line width to be 160$\kms$ for all  
sources, which is much larger than the value we use for most sources. 
Both \citet{McQuinn07} and \citet{Iliev08} assume the gas temperature in all
regions to be $10^4$K when computing the transmission of \lya emission.
Given these different assumptions used in our work and others, we perform
tests of the effects of initial line width and gas temperature on the \lya 
flux suppression in the $\exp(-\tau_\nu)$ model with a 
small simulation (box size of 25$\hMpc$ on a side). We calculate the 
distribution of the apparent-to-intrinsic \lya luminosity ratio for all 
sources with halo mass above $5\times 10^9\hMsun$. 

The solid curves in the left panel of Figure~\ref{fig:expmtau_test} show the
median and quartiles of the apparent-to-intrinsic \lya luminosity ratio 
as a function of the initial line width for the case using gas temperature in 
the reionization simulation. The results presented in \S~\ref{sec:statistics}
are based on the (default) value $\sigma_{\rm init}/\sigma_{\rm vir}=1$.
As expected, the luminosity ratio is sensitive to the initial line width. At 
$\sigma_{\rm init}/\sigma_{\rm vir}=2$ the median ratio is about one order of 
magnitude higher than that from our default case. 

The dotted curves in the left panel of Figure~\ref{fig:expmtau_test} 
correspond to the case with gas temperature set to $10^4$K in all regions. 
Changing the gas temperature leads to changes in the thermally averaged
cross-section for \lya scattering. As shown in the right panel of 
Figure~\ref{fig:expmtau_test}, the scattering cross-section 
is composed of a thermal (Gaussian) core and a Lorentzian wing dominated at 
small and large frequency offsets, respectively. The core is wider at higher
temperature. Since the temperature around a source in the reionization 
simulation can be higher than $10^4$K (Fig.~\ref{fig:halo_profile}), setting 
the temperature to $10^4$K would underestimate the scattering optical depth 
near the source and overestimate the transmitted flux. This explains why the 
luminosity ratio in the case with $T=10^4$K is higher than that with 
simulation temperature (left panel). For a quantitative
comparison, we note that \citet{Iliev08} present a case with 
$\sigma_{\rm init}/\sigma_{\rm vir}\simeq 2.2$ and $T=10^4$K (rightmost panel
of their Fig.25) and the median luminosity ratio is $\sim$0.33. Under the same 
assumptions, our $\exp(-\tau_\nu)$ model in Figure~\ref{fig:expmtau_test} gives
a median ratio of $\sim$0.25, in broad agreement with their result. The 
residual difference may be caused by the difference in the simulations
(e.g., hydrogen neutral fraction). 

To conclude, the outcome of the $\exp(-\tau_\nu)$ model depends on the 
initial \lya line width and gas temperature distribution and our 
$\exp(-\tau_\nu)$ model results are consistent with others if the same
assumptions are adopted.

{}


\begin{thebibliography}{}

\bibitem[Adams(1972)]{Adams72} 
Adams, T.~F.\ 1972, \apj, 174, 439 

\bibitem[Adelberger et al.(2003)]{Adelberger03} 
Adelberger, K.~L., Steidel, C.~C., Shapley, A.~E., \& Pettini, M.\ 2003, \apj, 
584, 45 

\bibitem[Ahn et al.(2000)]{Ahn00}
Ahn, S.-H., Lee, H.-W., \& Lee, H.~M.\ 2000, Journal of Korean Astronomical Society, 33, 29

\bibitem[Ahn et al.(2001)]{Ahn01} 
Ahn, S.-H., Lee, H.-W., \& Lee, H.M. 2001, ApJ, 554, 604

\bibitem[Ahn et al.(2002)]{Ahn02} 
Ahn, S.-H., Lee, H.-W., \& Lee, H.M. 2002, ApJ, 567, 922

\bibitem[Ando et al.(2006)]{Ando06} 
Ando, M., Ohta, K., Iwata, I., Akiyama, M., Aoki, K., \& Tamura, N.\ 2006, 
\apjl, 645, L9 

\bibitem[Ando et al.(2007)]{Ando07} 
Ando, M., Ohta, K., Iwata, I., Akiyama, M., Aoki, K., \& Tamura, N.\ 2007, 
\pasj, 59, 717 

\bibitem[Atek et al.(2008)]{Atek08} 
Atek, H., Kunth, D., Hayes, M., {\"O}stlin, G., \& Mas-Hesse, J.~M.\ 2008, 
\aap, 488, 491

\bibitem[Atek et al.(2009)]{Atek09} 
Atek, H., Schaerer, D., \& Kunth, D.\ 2009, \aap, 502, 791

\bibitem[Auer(1968)]{Auer68}  
Auer, L.~H.\ 1968, \apj, 153, 783 

\bibitem[Avery \& House(1968)]{Avery68} 
Avery, L.~W., \& House, L.~L.\ 1968, \apj, 152, 493 

\bibitem[Bouwens et al.(2003)]{Bouwens03} 
Bouwens, R.~J., et al.\ 2003, \apj, 595, 589 

\bibitem[Bouwens et al.(2006)]{Bouwens06}
Bouwens, R.~J., Illingworth, G.~D., Blakeslee, J.~P., \& Franx, M.\ 2006, \apj, 653, 53

\bibitem[Chabrier(2003)]{Chabrier03} 
Chabrier, G.\ 2003, \pasp, 115, 763 

\bibitem[Cowie \& Hu(1998)]{Cowie98} 
Cowie, L.~L., \& Hu, E.~M.\ 1998, \aj, 115, 1319 


\bibitem[Cuby et al.(2007)]{Cuby07} 
Cuby, J.-G., Hibon, P., Lidman, C., Le F{\`e}vre, O., Gilmozzi, R., Moorwood, A., \& van der Werf, P. 2007, A\&A, 461, 911

\bibitem[Dav{\'e}(2008)]{Dave08} 
Dav{\'e}, R.\ 2008, \mnras, 385, 147 

\bibitem[Dayal et al.(2008)]{Dayal08} 
Dayal, P., Ferrara, A., \& Gallerani, S.\ 2008, \mnras, 389, 1683 

\bibitem[Dayal et al.(2009)]{Dayal09} 
Dayal, P., Ferrara, A., Saro, A., Salvaterra, R., Borgani, S., 
\& Tornatore, L.\ 2009, \mnras, 400, 2000 

\bibitem[Dayal et al.(2010)]{Dayal10}
Dayal, P., Maselli, A., \& Ferrara, A.\ 2010, arXiv:1002.0839 

\bibitem[Dey et al.(1998)]{Dey98}
Dey, A., Spinrad, H., Stern, D., Graham, J.~R., \& Chaffee, F.~H.\ 1998, \apjl, 498, L93 

\bibitem[Dijkstra et al.(2006)]{Dijkstra06} 
Dijkstra, M., Haiman, Z., \& Spaans, M.\ 2006, \apj, 649, 14 

\bibitem[Dijkstra et al.(2007)]{Dijkstra07a} 
Dijkstra, M., Lidz, A., \& Wyithe, J.~S.~B.\ 2007, \mnras, 377, 1175 

\bibitem[Dijkstra et al.(2007)]{Dijkstra07b} 
Dijkstra, M., Wyithe, J.~S.~B., \& Haiman, Z.\ 2007, \mnras, 379, 253 

\bibitem[Dijkstra et al.(2006a)]{Dijkstra06a} 
Dijkstra, M., Haiman, Z., \& Spaans, M. 2006a, ApJ, 649, 14

\bibitem[Dijkstra et al.(2006b)]{Dijkstra06b} 
Dijkstra, M., Haiman, Z., \& Spaans, M. 2006b, ApJ, 649, 37

\bibitem[Dow-Hygelund et al.(2007)]{Dow-Hygelund07}
Dow-Hygelund, C.~C., et al.\ 2007, \apj, 660, 47 

\bibitem[Dunkley et al.(2009)]{Dunkley09} 
Dunkley, J., et al.\ 2009, \apjs, 180, 306 

\bibitem[Eisenstein et al.(2005)]{Eisenstein05} 
Eisenstein, D.~J., et al.\ 2005, \apj, 633, 560 

\bibitem[Franx et al.(1997)]{Franx97} 
Franx, M., Illingworth, G.~D., Kelson, D.~D., van Dokkum, P.~G., 
\& Tran, K.-V.\ 1997, \apjl, 486, L75 

\bibitem[Furlanetto et al.(2005)]{Furlanetto05} 
Furlanetto, S.~R., Schaye, J., Springel, V., \& Hernquist, L.\ 2005, \apj, 
622, 7 

\bibitem[Furlanetto et al.(2006)]{Furlanetto06} 
Furlanetto, S.R., Zaldarriaga, M., \& Hernquist, L. 2006, MNRAS, 365, 1012

\bibitem[Giavalisco et al.(1996)]{Giavalisco96}
Giavalisco, M., Koratkar, A., \& Calzetti, D.\ 1996, \apj, 466, 831


\bibitem[Haiman \& Cen(2005)]{Haiman05} 
Haiman, Z., \& Cen, R. 2005,  ApJ, 623, 627

\bibitem[Haiman \& Spaans(1999)]{Haiman99} 
Haiman, Z., \& Spaans, M. 1999, ApJ, 518, 138

\bibitem[Hansen \& Oh(2006)]{Hansen06}
Hansen, M., \& Oh, S.~P.\ 2006, \mnras, 367, 979

\bibitem[Harrington(1973)]{Harrington73}
Harrington, J.~P.\ 1973, \mnras, 162, 43

\bibitem[Harrington(1974)]{Harrington74}
Harrington, J.~P.\ 1974, \mnras, 166, 373

\bibitem[Hill et al.(2008)]{Hill08} 
Hill, G.~J., et al.\ 2008, Astronomical Society of the Pacific Conference 
Series, 399, 115 

\bibitem[Horton et al.(2004)]{Horton04} 
Horton, A., Parry, I., Bland-Hawthorn, J., Cianci, S., King, D., McMahon, R., \& Medlen, S. 2004, Ground-based Instrumentation for Astronomy, ed. A.F.~M.~Moorwood and I.  Masanori, Proceedings of the SPIE, 5492, p1022

\bibitem[Hu \& McMahon(1996)]{Hu96} 
Hu, E., \& McMahon, R.~G.\ 1996, \nat, 382, 281 

\bibitem[Hu et al.(1998)]{Hu98} 
Hu, E. M., Cowie, L. L., \& McMahon, R. G. 1998, ApJ, 502, L99 

\bibitem[Hu et al.(1999)]{Hu99} 
Hu, E.~M., McMahon, R.~G., \& Cowie, L.~L.\ 1999, \apjl, 522, L9 

\bibitem[Hu et al.(2002)]{Hu02} 
Hu, E. M., Cowie, L. L., McMahon, R. G., Capak, P., Iwamuro, F., Kneib, J.-P., Maihara, T., \& Motohara, K. 2002, ApJ, 568, L75 (erratum 576, L99) 

\bibitem[Hu et al.(2004)]{Hu04} 
Hu, E. M., Cowie, L. L., Capak, P., McMahon, R. G., Hayashino, T., \& Komiyama, Y. 2004, AJ, 127, 563 

\bibitem[Hu et al.(2005)]{Hu05} 
Hu, E. M., Cowie, L. L., Capak, P., \& Kakazu, Y. 2005, IAU Colloq. 199, Probing Galaxies through Quasar Absorption Lines, ed. P. Williams, C.-G. Shu, \& B. Menard (Cambridge: Cambridge Univ. Press), p363 

\bibitem[Hu et al.(2006)]{Hu06} 
Hu, E.M., Cowie, L.L., \& Kakazu, Y. 2006, The Universe at $z>6$, 26th meeting of the IAU, (Prague, Czech Republic), JD07, 10

\bibitem[Hummer(1962)]{Hummer62}  
Hummer, D.~G.\ 1962, \mnras, 125, 21 

\bibitem[Iliev et al.(2008)]{Iliev08}
Iliev, I.~T., Shapiro, P.~R., McDonald, P., Mellema, G., \& Pen, U.-L.\ 2008, 
\mnras, 391, 63

\bibitem[Iye et al.(2006)]{Iye06}
Iye, M., et al. 2006, Nature, 443, 186

\bibitem[Kobayashi et al.(2010)]{Kabayashi10}
Kobayashi, M.~A.~R., Totani, T., \& Nagashima, M.\ 2010, \apj, 708, 1119 


\bibitem[Kajino et al.(2009)]{Kajino09} 
Kajino, H., et al.\ 2009, \apj, 704, 117 

\bibitem[Kashikawa et al.(2006)]{Kashikawa06}
Kashikawa, N., et al. 2006, ApJ, 648, 7

\bibitem[Keel(2005)]{Keel05} 
Keel, W.~C.\ 2005, \aj, 129, 1863

\bibitem[Kollmeier et al.(2010)]{Kollmeier10}
Kollmeier, J.~A., Zheng, Z., Dav{\'e}, R., Gould, A., Katz, N.,
Miralda-Escud{\'e}, J., \& Weinberg, D.~H.\ 2010, \apj, 708, 1048 

\bibitem[Laursen \& Sommer-Larsen(2007)]{Laursen07} 
Laursen, P., \& Sommer-Larsen, J.\ 2007, \apjl, 657, L69 

\bibitem[Laursen et al.(2009)]{Laursen09}
Laursen, P., Razoumov, A.~O., \& Sommer-Larsen, J.\ 2009, \apj, 696, 853

\bibitem[Loeb \& Rybicki(1999)]{Loeb99} 
Loeb, A., \& Rybicki, G.~B.\ 1999, \apj, 524, 527 

\bibitem[Madau et al.(1998)]{Madau98}
Madau, P., Pozzetti, L., \& Dickinson, M.\ 1998, \apj, 498, 106

\bibitem[Malhotra \& Rhoads(2004)]{Malhotra04} 
Malhotra, S., \& Rhoads, J. 2004, ApJ, 617, L5

\bibitem[Mao et al.(2007)]{Mao07}
Mao, J., Lapi, A., Granato, G.~L., de Zotti, G., \& Danese, L.\ 2007, \apj, 
667, 655 

\bibitem[McQuinn et al.(2007)]{McQuinn07} 
McQuinn, M., Hernquist, L., Zaldarriaga, M., \& Dutta, S.\ 2007, \mnras, 381, 
75 

\bibitem[Mesinger \& Furlanetto(2008)]{Mesinger08} 
Mesinger, A., \& Furlanetto, S.R. 2008, MNRAS, 386, 1990

\bibitem[Miralda-Escud\'e \& Rees(1998)]{Rees98} 
Miralda-Escud\'e, J., \& Rees, M.J. 1998, ApJ, 497, 21

\bibitem[Miralda-Escud\'e(1998)]{Miralda98} 
Miralda-Escud\'e, J. 1998, ApJ, 501, 15

\bibitem[Nagamine et al.(2008)]{Nagamine08} 
Nagamine, K., Ouchi, M., Springel, V., \& Hernquist, L.\ 2008, arXiv:0802.0228 

\bibitem[Neufeld(1990)]{Neufeld90}
Neufeld, D.~A.\ 1990, \apj, 350, 216

\bibitem[Neufeld(1991)]{Neufeld91}
Neufeld, D.~A.\ 1991, \apjl, 370, L85

\bibitem[Nilsson et al.(2007)]{Nilsson07} 
Nilsson, K.K., Orsi, A., Lacey, C.G., Baugh, C.M., \& Thommes, E. 2007, A\& A, 474, 385

\bibitem[Nilsson et al.(2009)]{Nilsson09} 
Nilsson, K.~K., Moeller-Nilsson, O., Moeller, P., Fynbo, J.~P.~U., 
\& Shapley, A.~E.\ 2009, \mnras, 400, 232

\bibitem[Osterbrock(1989)]{Osterbrock89} 
Osterbrock, D.~E.\ 1989, Astrophysics of Gaseous Nebulae and Active Galactic Nuclei (Mill Valley, CA: Univ. Sci.)

\bibitem[Ota et al.(2008)]{Ota08} 
Ota, K., et al.\ 2008, ApJ, 677, 12


\bibitem[Ouchi et al.(2007)]{Ouchi07} 
Ouchi, M., Tokoku, C., Shimasaku, K., \& Ichikawa, T. 2007, ASP Conf.~Ser., ed. N. Metcalfe and T. Shanks, p47 

\bibitem[Ouchi et al.(2008)]{Ouchi08} 
Ouchi, M., et al.\ 2008, \apjs, 176, 301

\bibitem[Ouchi et al.(2009)]{Ouchi09} 
Ouchi, M., et al.\ 2009, \apj, 706, 1136 

\bibitem[Partridge \& Peebles(1967)]{Partridge67}
Partridge, R.B., \& Peebles, P.J.E. 1967, ApJ, 147, 868

\bibitem[Pentericci et al.(2009)]{Pentericci09} 
Pentericci, L., Grazian, A., Fontana, A., Castellano, M., Giallongo, E., 
Salimbeni, S., \& Santini, P.\ 2009, \aap, 494, 553 

\bibitem[Pierleoni et al.(2009)]{Pierleoni09}
Pierleoni, M., Maselli, A., \& Ciardi, B.\ 2009, \mnras, 393, 872

\bibitem[Rhoads et al.(2003)]{Rhoads03} 
Rhoads, J.E., Dey, A., Malhotra, S., Stern, D., Spinrad, H., Jannuzi, B.T., Dawson, S., Brown, M.J.I., \& Landes, E. 2003, AJ, 125, 1006

\bibitem[Salpeter(1955)]{Salpeter55} 
Salpeter, E.~E.\ 1955, \apj, 121, 161

\bibitem[Samui et al.(2009)]{Samui09} 
Samui, S., Srianand, R., \& Subramanian, K.\ 2009, \mnras, 1183 

\bibitem[Santos(2004)]{Santos04}
Santos, M.~R.\ 2004, \mnras, 349, 1137 

\bibitem[Schaerer(2003)]{Schaerer03} 
Schaerer, D.\ 2003, \aap, 397, 527 

\bibitem[Shapley et al.(2003)]{Shapley03} 
Shapley, A.~E., Steidel, C.~C., Pettini, M., \& Adelberger, K.~L.\ 2003, 
\apj, 588, 65 

\bibitem[Sheth \& Tormen(1999)]{Sheth99}
Sheth, R.\ K.\ \& Tormen, G.\ 1999, \mnras, 308, 119

\bibitem[Shimasaku et al.(2006)]{Shimasaku06} 
Shimasaku, K., et al.\ 2006, \pasj, 58, 313 

\bibitem[Shioya et al.(2009)]{Shioya09} 
Shioya, Y., et al.\ 2009, \apj, 696, 546 

\bibitem[Stanway et al.(2007)]{Stanway07} 
Stanway, E.~R., et al.\ 2007, \mnras, 376, 727 

\bibitem[Stark et al.(2007a)]{Stark07a} 
Stark, D.~P., Ellis, R.~S., Richard, J., Kneib, J.-P., Smith, G.~P., 
\& Santos, M.~R.\ 2007, \apj, 663, 10 

\bibitem[Stark et al.(2007b)]{Stark07b} 
Stark, D.~P., Loeb, A., \& Ellis, R.~S.\ 2007, \apj, 668, 627 

\bibitem[Stern et al.(2005)]{Stern05} 
Stern, D., Yost, S. A., Eckart, M. E., Harrison, F. A., Helfand, D. J., Djorgovski, S. G., Malhotra, S., \& Rhoads, J. E. 2005, ApJ, 619, 12 

\bibitem[Taniguchi et al.(2009)]{Taniguchi09}
Taniguchi, Y., et al.\ 2009, \apj, 701, 915 

\bibitem[Tasitsiomi(2006)]{Tasitsiomi06} 
Tasitsiomi, A.\ 2006, \apj, 645, 792 

\bibitem[Tilvi et al.(2009)]{Tilvi09} 
Tilvi, V., Malhotra, S., Rhoads, J.~E., Scannapieco, E., Thacker, R.~J., 
Iliev, I.~T., \& Mellema, G.\ 2009, \apj, 704, 724 

\bibitem[Trac \& Pen(2004)]{Trac04} 
Trac, H., \& Pen, U.-L.\ 2004, New Astronomy, 9, 443 

\bibitem[Trac \& Pen(2006)]{Trac06} 
Trac, H., \& Pen, U.-L.\ 2006, New Astronomy, 11, 273 

\bibitem[Trac \& Cen(2007)]{Trac07} 
Trac, H., \& Cen, R.\ 2007, \apj, 671, 1 

\bibitem[Trac et al.(2008)]{Trac08} 
Trac, H., Cen, R., \& Loeb, A.\ 2008, \apjl, 689, L81 

\bibitem[van Dokkum(2008)]{vanDokkum08} 
van Dokkum, P.~G.\ 2008, \apj, 674, 29 

\bibitem[Vanzella et al.(2009)]{Vanzella09} 
Vanzella, E., et al.\ 2009, \apj, 695, 1163 

\bibitem[Verhamme et al.(2006)]{Verhamme06}
Verhamme, A., Schaerer, D., \& Maselli, A.\ 2006, \aap, 460, 397

\bibitem[Willis et al.(2008)]{Willis08} 
Willis, J.P., Courbin, F., Kneib, J.-P., \& Minniti, D. 2008, MNRAS, 384, 1039

\bibitem[Wyithe \& Cen(2007)]{Wyithe07} 
Wyithe, J.~S.~B., \& Cen, R.\ 2007, \apj, 659, 890 


\bibitem[Zheng \& Miralda-Escud{\'e}(2002)]{Zheng02} 
Zheng, Z., \& Miralda-Escud{\'e}, J.\ 2002, \apj, 578, 33 

\end{thebibliography}
\end{document}